\documentclass[11pt]{article}%
\usepackage{amssymb}
\usepackage[singlespacing]{setspace}
\usepackage{amsmath}
\usepackage{amsfonts}
\usepackage{graphicx}
\usepackage{cite}%
\allowdisplaybreaks
\begin{document}

\author{\vspace{0.16in}Hartmut Wachter\thanks{e-mail:
Hartmut.Wachter@physik.uni-muenchen.de}\\Max-Planck-Institute\\for Mathematics in the Sciences\\Inselstr. 22, D-04103 Leipzig\\\hspace{0.4in}\\Arnold-Sommerfeld-Center\\Ludwig-Maximilians-Universit\"{a}t\\Theresienstr. 37, D-80333 M\"{u}nchen}
\title{Non-relativistic Schr\"{o}dinger theory on q-deformed quantum spaces III\\{\small Scattering theory}}
\date{}
\maketitle

\begin{abstract}
\noindent This is the third part of a paper about non-relativistic
Schr\"{o}dinger theory on q-deformed quantum spaces like the braided line or
the three-dimensional q-deformed Euclidean space. Propagators for the free
q-deformed particle are derived and their basic properties are discussed. A
time-dependent formulation of scattering is proposed. In this respect,
q-analogs of the Lippmann-Schwinger equation are given. Expressions for their
iterative solutions are written down. It is shown how to calculate S-matrices
and transition probabilities. Furthermore, attention is focused on the
question what becomes of unitarity of S-matrices in a q-deformed setting. The
examinations are concluded by a discussion of the interaction picture and its
relation to scattering processes.\newpage

\end{abstract}
\tableofcontents

\section{Introduction}

It is an old idea to formulate quantum field theories on a space-time lattice,
since it should lead to a natural cut-off in momentum space \cite{Schw,Heis}.
In the literature one can find\ several attempts to attack this problem (see
for example Refs. \cite{Sny47, Yan47, Fli48, Hill55, Das60, Gol63}). A more
recent but very promising approach to discretize space-time is based on the
theory of quantum groups and quantum spaces \cite{Ku83, Wor87, Dri85, Jim85,
Drin86, RFT90, Tak90, WZ91, CSW91, Song92, OSWZ92, Maj93-2, FLW96, Wes97,
CW98, GKP96, MajReg, Oec99, Maj95, KS97, ChDe96, Man88, Maj94-10, Maj93-Int,
Wess00, Maj95star, CSSW90, PW90, SWZ91, Maj91, LWW97}. In our previous work we
tried to develop the concepts of this theory in a way that allows their
application to quantum theory \cite{WW01, BW01, Wac02, Wac03, Wac04, Wac05,
MSW04, SW04, qAn, Qkin1, Qkin2}.

This is the third and last part of an article which continues our former
reasonings on physical aspects of the theory of quantum groups and quantum
spaces. More concretely, the article is devoted to a non-relativistic
Schr\"{o}dinger theory on q-deformed quantum spaces as the braided line or the
three-dimensional q-deformed Euclidean space. Before we summarize the content
of its third part, let us briefly recall what we have already done in part I
and II.

In part I we first presented the algebraic structure of the quantum spaces
under consideration. Then we adapted our reasonings about q-deformed versions
of analysis to the braided line and the three-dimensional q-deformed Euclidean
space. Finally, we showed that this\ mathematical framework is compatible with
basic concepts of quantum dynamics. Especially, we saw that the time-evolution
operators for quantum spaces are of the same form as in the undeformed case.
Furthermore, we gave q-analogs of the Schr\"{o}dinger equation and the
Heisenberg equations of motion.

Part II of the paper applies the reasonings of part I\ to the free
non-relativistic particle. Especially, we considered q-analogs of momentum
eigenfunctions and discussed their completeness, orthonormality, and
dependence from time. In addition to this, we dealt with some more general
aspects of quantum theory, i.e. the theorem of Ehrenfest and the conservation
of probability. It was shown how these notions carry over to our q-deformed spaces.

Now, we come to the content of part III, which\ concludes our examinations of
a q-deformed analog of non-relativistic Schr\"{o}dinger theory. This part
provides us with basic concepts of non-relativistic scattering theory on
q-deformed quantum spaces. Towards this end, we reconsider the problem of
time-evolution in quantum mechanics and treat it from the point of view of
propagator theory. This task will be done in Sec.\thinspace\ref{FrePro}. In
doing so, we obtain q-deformed expressions for\ the propagator of a free
non-relativistic particle. Furthermore, we give a collection of the most
important properties of these q-deformed propagators. After that we should be
prepared to establish a q-deformed version of non-relativistic scattering
theory. Section \ref{ScaThe} covers this subject. More concretely, we first
derive q-analogs of the Lippmann-Schwinger equation and solve them
iteratively. This leads us to expansions that are relevant in perturbation
theory. Furthermore, we introduce Green's functions for a particle interacting
with a potential, write down their perturbation expansions, and discuss their
basic properties. The matrix elements of these Green's functions can be used
to calculate S-matrices and transition probabilities. Finally, we concern
ourselves with the question what becomes of unitarity of S-matrices in a
q-deformed setting. In Sec.\thinspace\ref{IntPic} we revisit time-dependent
scattering theory and treat it from the point of view provided by the
interaction picture. Section \ref{SecCon} closes our considerations by a short conclusion.

The line of our reasonings is very similar to that in the undeformed case.
However, there is one remarkable difference between a q-deformed theory and
its undeformed limit, since in a q-deformed theory we have to distinguish
different geometries. The reason for this lies in the fact that the braided
tensor category in which the expressions of our theory live is not uniquely
determined. This becomes more clear, if one realizes that each braided
category is characterized by a so-called braiding mapping $\Psi$. The inverse
$\Psi^{-1}$ gives an equally good braiding, which leads to a second braided
category being different from the first one. This observation is reflected in
the occurrence of two\ differential calculi, different types of
q-exponentials, q-integrals and so on. In this manner each braided category
implies its own q-geometry, so we can write down different q-analogs of
well-known physical laws.

Lastly, it should be noted that we assume the reader to be familiar with the
results and conventions of part I and II. In this respect, it would be helpful
to have an idea of the reasonings in Sec.\thinspace3.1 or 3.2 of part I.
Furthermore, we recommend to have a look at Sec.\thinspace2.2 and 2.3 of part II.

\section{Propagators of the free particle \label{FrePro}}

Once the wave function of a quantum system is known at a certain time the
time-evolution operator enables us to find the wave function at any later
time. However, there is another way to solve the time-evolution problem. It
requires to know the so-called propagators. In this section we give
expressions for the propagator of the q-deformed non-relativistic free
particle and derive some of their basic properties.

We start from the expansions of wave functions in terms of plane waves. There
are different q-geometries and each geometry leads to its own expansion. In
part II of the article we found that%
\begin{align}
(\phi_{1})_{m}^{\prime}(x^{i})  &  =\frac{\kappa^{n}}{(\text{vol}_{1})^{1/2}%
}\int_{-\infty}^{+\infty}d_{1}^{n}p\,(c_{1})_{\kappa p}^{\prime}\overset
{p|x}{\odot}_{\hspace{-0.01in}R}(u_{\bar{R},L})_{\ominus_{L}p,m}%
(x^{i}),\nonumber\\
(\phi_{2})_{m}(x^{i})  &  =\frac{\kappa^{-n}}{(\text{vol}_{2})^{1/2}}%
\int_{-\infty}^{+\infty}d_{2}^{n}p\,(\bar{u}_{R,\bar{L}})_{\ominus_{R}%
p,m}(x^{i})\overset{x|p}{\odot}_{\hspace{-0.01in}L}(c_{2})_{\kappa^{-1}%
p},\label{ExpWav1}\\[0.1in]
(\phi_{1}^{\ast})_{m}(x^{i})  &  =(\text{vol}_{1})^{1/2}\int_{-\infty
}^{+\infty}d_{1}^{n}p\,(u_{\bar{R},L})_{p,m}(x^{i})\overset{p}{\circledast
}(c_{1}^{\ast})_{\kappa^{-1}p},\nonumber\\
(\phi_{2}^{\ast})_{m}^{\prime}(x^{i})  &  =(\text{vol}_{2})^{1/2}\int
_{-\infty}^{+\infty}d_{2}^{n}p\,(c_{2}^{\ast})_{\kappa p}^{\prime}\overset
{p}{\circledast}(\bar{u}_{R,\bar{L}})_{p,m}(x^{i}), \label{ExpWav2}%
\end{align}
where the expansion coefficients can be calculated from the formulae%
\begin{align}
(c_{1})_{p}^{\prime}  &  =(\text{vol}_{1})^{1/2}\int\nolimits_{-\infty
}^{+\infty}d_{1}^{n}x\,(\phi_{1})_{m}^{\prime}(x^{i})\overset{x}{\circledast
}(u_{\bar{R},L})_{p,m}(x^{A},-q^{\zeta}t_{x}),\nonumber\\
(c_{2})_{p}  &  =(\text{vol}_{2})^{1/2}\int\nolimits_{-\infty}^{+\infty}%
d_{2}^{n}x\,(\bar{u}_{R,\bar{L}})_{p,m}(x^{A},-q^{-\zeta}t_{x})\overset
{x}{\circledast}(\phi_{2})_{m}(x^{i}),
\end{align}
and%
\begin{align}
(c_{1}^{\ast})_{p}  &  =\frac{1}{(\text{vol}_{1})^{1/2}}\int\nolimits_{-\infty
}^{+\infty}d_{1}^{n}x\,(u_{\bar{R},L})_{\ominus_{\bar{R}}p,m}(x^{A}%
,-q^{-\zeta}t_{x})\overset{p|x}{\odot}_{\hspace{-0.01in}\bar{L}}(\phi
_{1}^{\ast})_{m}(x^{i}),\nonumber\\
(c_{2}^{\ast})_{p}^{\prime}  &  =\frac{1}{(\text{vol}_{2})^{1/2}}%
\int\nolimits_{-\infty}^{+\infty}d_{2}^{n}x\,(\phi_{2}^{\ast})_{m}^{\prime
}(x^{i})\overset{x|p}{\odot}_{\hspace{-0.01in}\bar{R}}(\bar{u}_{R,\bar{L}%
})_{\ominus_{\bar{L}}p,m}(x^{A},-q^{\zeta}t_{x}).
\end{align}

Let us recall that for the quantum spaces under consideration the values of
$\kappa$ and $\zeta$ are determined as follows:

\begin{enumerate}
\item[(i)] (braided line) $\kappa=q,$ $\zeta=-1,$

\item[(ii)] (q-deformed Euclidean space in three dimensions) $\kappa=q^{6}$,
$\zeta=2.$
\end{enumerate}

\noindent The q-analogs of plane-waves are related to q-deformed exponentials
\cite{Wac03, Schir94, Maj93-5} by%
\begin{align}
(u_{\bar{R},L})_{p,m}(x^{i}) &  =(\text{vol}_{1})^{-1/2}\exp(x^{i}%
|\text{i}^{-1}p_{j})_{\bar{R},L}\big |_{p_{0}=p^{2}/(2m)\overset
{p}{\circledast}},\nonumber\\
(u_{R,\bar{L}})_{p,m}(x^{i}) &  =(\text{vol}_{2})^{-1/2}\exp(x^{i}%
|\text{i}^{-1}p_{j})_{R,\bar{L}}\big |_{p_{0}=p^{2}/(2m)\overset
{p}{\circledast}},\\[0.1in]
(\bar{u}_{\bar{R},L})_{p,m}(x^{i}) &  =(\text{vol}_{1})^{-1/2}\exp
(\text{i}^{-1}p_{j}|x^{i})_{\bar{R},L}\big |_{p_{0}=p^{2}/(2m)\overset
{p}{\circledast}},\nonumber\\
(\bar{u}_{R,\bar{L}})_{p,m}(x^{i}) &  =(\text{vol}_{2})^{-1/2}\exp
(\text{i}^{-1}p_{j}|x^{i})_{R,\bar{L}}\big |_{p_{0}=p^{2}/(2m)\overset
{p}{\circledast}}.
\end{align}
The symbols $\odot_{\hspace{-0.01in}\gamma},$ $\gamma\in\{L,\bar{L},R,\bar
{R}\},$ and $\circledast$ respectively denote braided products and star
multiplication. Notice that braided products represent realizations of
braiding mappings \cite{Wac05}. The symbol on top of a braided product
indicates the tensor factors being involved in the braiding. It should also be
mentioned that the volume elements are given by%
\begin{align}
\text{vol}_{1} &  \equiv\int_{-\infty}^{+\infty}d_{1}^{n}x\int_{-\infty
}^{+\infty}d_{1}^{n}p\exp(x^{i}|\text{i}^{-1}p_{j})_{\bar{R},L}\big |_{x^{0}%
=0}\nonumber\\
&  =\int_{-\infty}^{+\infty}d_{1}^{n}p\int_{-\infty}^{+\infty}d_{1}^{n}%
x\exp(\text{i}^{-1}p_{i}|x^{j})_{\bar{R},L}\big |_{x^{0}=0},\\[0.1in]
\text{vol}_{2} &  \equiv\int_{-\infty}^{+\infty}d_{2}^{n}x\int_{-\infty
}^{+\infty}d_{2}^{n}p\exp(x^{i}|\text{i}^{-1}p_{j})_{R,\bar{L}}\big |_{x^{0}%
=0}\nonumber\\
&  =\int\nolimits_{-\infty}^{+\infty}d_{2}^{n}p\int_{-\infty}^{+\infty}%
d_{2}^{n}x\exp(\text{i}^{-1}p_{i}|x^{j})_{R,\bar{L}}\big |_{x^{0}=0}.
\end{align}
Expressions for calculating q-integrals over the whole position or momentum
space were listed in part I of the paper [cf. Sec.\thinspace5 of part I].

As we already know from part I we have to distinguish different q-geometries.
In this article, however, we restrict attention to certain geometries, only,
since the expressions for the other geometries can be obtained from our
results by applying the substitutions%
\begin{gather}
L\leftrightarrow\bar{L},\quad R\leftrightarrow\bar{R},\quad\kappa
\leftrightarrow\kappa^{-1},\quad q\leftrightarrow q^{-1},\nonumber\\
1\text{ (as label)}\leftrightarrow2\text{ (as label)},\quad\partial
\leftrightarrow\hat{\partial},\quad\triangleright\leftrightarrow
\bar{\triangleright},\quad\triangleleft\leftrightarrow\bar{\triangleleft}.
\label{SubGeo}%
\end{gather}
These correspondences are an example for so-called crossing symmetries, which
are typical for q-deformation (see also Ref. \cite{qAn}).

Now, we come to the derivation of free-particle propagators. Inserting the
expressions for the expansion coefficients into the expansions of
(\ref{ExpWav1}) and (\ref{ExpWav2}) leads us to integral operators that act on
the initial wave functions to yield the final wave functions:%
\begin{align}
(\phi_{1})_{m}^{\prime}(x^{i})  &  =\int\nolimits_{-\infty}^{+\infty}d_{1}%
^{n}y\,(\phi_{1})_{m}^{\prime}(y^{j})\overset{y}{\circledast}(K_{1}%
)_{m}^{\prime}(y^{k},x^{i}),\nonumber\\
(\phi_{2})_{m}(x^{i})  &  =\int\nolimits_{-\infty}^{+\infty}d_{2}^{n}%
y\,(K_{2})_{m}(x^{i},y^{j})\overset{y}{\circledast}(\phi_{2})_{m}%
(y^{k}),\\[0.1in]
(\phi_{1}^{\ast})_{m}(x^{i})  &  =\int\nolimits_{-\infty}^{+\infty}d_{1}%
^{n}y\,(K_{1}^{\ast})_{m}(x^{i},y^{j})\overset{y}{\circledast}(\phi_{1}^{\ast
})_{m}(\kappa y^{A},t_{y}),\nonumber\\
(\phi_{2}^{\ast})_{m}^{\prime}(x^{i})  &  =\int\nolimits_{-\infty}^{+\infty
}d_{2}^{n}y\,(\phi_{2}^{\ast})_{m}^{\prime}(\kappa^{-1}y^{A},t_{y})\overset
{y}{\circledast}(K_{2}^{\ast})_{m}^{\prime}(y^{k},x^{i}).
\end{align}
The kernels of these integral operators are known as propagators and take the
form%
\begin{align}
(K_{1})_{m}^{\prime}(y^{i},x^{j})  &  =\kappa^{n}\int_{-\infty}^{+\infty}%
d_{1}^{n}p\,(u_{\bar{R},L})_{\kappa p,m}(y^{A},-q^{\zeta}\kappa^{-2}%
t_{y})\nonumber\\
&  \hspace{1.2in}\overset{p|x}{\odot}_{\hspace{-0.01in}R}(u_{\bar{R}%
,L})_{\ominus_{L}p,m}(x^{B},t_{x}),\\[0.08in]
(K_{2})_{m}(x^{i},y^{j})  &  =\kappa^{-n}\int_{-\infty}^{+\infty}d_{2}%
^{n}p\,(\bar{u}_{R,\bar{L}})_{\ominus_{R}p,m}(x^{A},t_{x})\nonumber\\
&  \hspace{1.2in}\overset{x|p}{\odot}_{\hspace{-0.01in}L}(\bar{u}_{R,\bar{L}%
})_{\kappa^{-1}p,m}(y^{B},-q^{-\zeta}\kappa^{2}t_{y}),\\[0.08in]
(K_{1}^{\ast})_{m}(x^{i},y^{j})  &  =\kappa^{n}\int_{-\infty}^{+\infty}%
d_{1}^{n}p\,(u_{\bar{R},L})_{p,m}(x^{A},t_{x})\nonumber\\
&  \hspace{1.2in}\overset{p|x}{\odot}_{\hspace{-0.01in}\bar{L}}(u_{\bar{R}%
,L})_{\ominus_{\bar{R}}p,m}(y^{B},-q^{-\zeta}t_{y}),\\[0.08in]
(K_{2}^{\ast})_{m}^{\prime}(y^{i},x^{j})  &  =\kappa^{-n}\int_{-\infty
}^{+\infty}d_{2}^{n}p\,(\bar{u}_{R,\bar{L}})_{\ominus_{R}p,m}(y^{A},-q^{\zeta
}t_{y})\nonumber\\
&  \hspace{1.2in}\overset{x|p}{\odot}_{\hspace{-0.01in}\bar{R}}(\bar
{u}_{R,\bar{L}})_{p,m}(x^{B},t_{x}).
\end{align}
At this point it should be mentioned that we take the convention from part I
and II that capital letters like $A,$ $B,$ etc. denote indices of space
coordinates, only, i.e., for example, $x^{i}=(x^{A},x^{0})=(x^{A},t).$

We have to impose on the propagators the causality requirement, i.e. the wave
function at time $t$ can not be influenced by the wave functions at times
$t^{\prime}>t.$ This leads us to the retarded Green's functions%
\begin{align}
(K_{1})_{m^{+}}^{\prime}(y^{i},x^{j})  &  \equiv\theta(t_{x}-t_{y})(K_{1}%
)_{m}^{\prime}(y^{A},t_{y};x^{B},t_{x}),\nonumber\\
(K_{2})_{m^{+}}(x^{i},y^{j})  &  \equiv\theta(t_{x}-t_{y})(K_{2})_{m}%
(x^{A},-t_{x};y^{B},-t_{y}),\\[0.1in]
(K_{1}^{\ast})_{m^{+}}(x^{i},y^{j})  &  \equiv\theta(t_{x}-t_{y})(K_{1}^{\ast
})_{m}(x^{A},t_{x};y^{B},t_{y}),\nonumber\\
(K_{2}^{\ast})_{m^{+}}^{\prime}(y^{i},x^{j})  &  \equiv\theta(t_{x}%
-t_{y})(K_{2}^{\ast})_{m}^{\prime}(y^{A},-t_{y};x^{B},-t_{x}),
\end{align}
where $\theta(t)$ stands for the Heaviside function%
\begin{equation}
\theta(t)=%
\begin{cases}
1 & \text{if }t\geq0,\\
0 & \text{otherwise}.
\end{cases}
\end{equation}
However, nothing prevents us from dealing with advanced Green's functions.
They should be introduced in a way\ that it holds%
\begin{align}
(K_{1})_{m^{-}}^{\prime}(y^{i},x^{j})  &  \equiv(K_{1})_{m^{+}}^{\prime}%
(y^{A},-t_{y};x^{B},-t_{x}),\nonumber\\
(K_{2})_{m^{-}}(x^{i},y^{j})  &  \equiv(K_{2})_{m^{+}}(y^{A},-t_{y}%
;x^{B},-t_{x}),\\[0.1in]
(K_{1}^{\ast})_{m^{-}}(x^{i},y^{j})  &  \equiv(K_{1}^{\ast})_{m^{+}}%
(y^{A},-t_{y};x^{B},-t_{x}),\nonumber\\
(K_{2}^{\ast})_{m^{-}}^{\prime}(y^{i},x^{j})  &  \equiv(K_{2}^{\ast})_{m^{+}%
}^{\prime}(y^{A},-t_{y};x^{B},-t_{x}).
\end{align}

Next, we wish to check that these Green's functions are solutions to
Schr\"{o}dinger equations with q-deformed delta functions as potential. This
assertion follows from the same arguments as in the undeformed case:%
\begin{align}
\text{i}\partial_{0}\overset{t_{x}}{\triangleright}(K_{1})_{m^{+}}^{\prime
}(y^{i},x^{j})=  &  \,\,(\text{i}\partial_{0}\overset{t_{x}}{\triangleright
}\theta(t_{x}-t_{y}))(K_{1})_{m}^{\prime}(y^{i},x^{j})\nonumber\\
&  \,\,+\,\theta(t_{x}-t_{y})(\text{i}\partial_{0}\overset{t_{x}%
}{\triangleright}(K_{1})_{m}^{\prime}(y^{i},x^{j}))\nonumber\\
=  &  \,\,\text{i}\delta(t_{x}-t_{y})(K_{1})_{m}^{\prime}(y^{i},x^{j}%
)\nonumber\\
&  \,\,+\,\theta(t_{x}-t_{y})(H_{0}\overset{x}{\triangleright}(K_{1}%
)_{m}^{\prime}(y^{i},x^{j}))\nonumber\\
=  &  \,\,\text{i}\kappa^{n}(\text{vol}_{1})^{-1}\delta(t_{x}-t_{y}%
)\,\delta_{1}^{n}(\kappa y^{A}\oplus_{\bar{R}}(\ominus_{\bar{R}}%
\,x^{B}))\nonumber\\
&  \,\,\,+\,H_{0}\overset{x}{\triangleright}(K_{1})_{m^{+}}^{\prime}%
(y^{i},x^{j}). \label{SchrGlGre}%
\end{align}
The first step uses the definition of the retarded Green's function and the
Leibniz rule for the time derivative. For the second step we make use of the
fact that the time derivative of the Heaviside function is given by the
classical delta function. Furthermore, we apply%
\begin{align}
&  \text{i}\partial_{0}\overset{t_{x}}{\triangleright}(K_{1})_{m}^{\prime
}(y^{i},x^{j})\nonumber\\
&  \quad=\,\kappa^{n}\int_{-\infty}^{+\infty}d_{1}^{n}p\,(u_{\bar{R}%
,L})_{\kappa p,m}(y^{A},-q^{\zeta}\kappa^{-2}t_{y})\overset{p|x}{\odot
}_{\hspace{-0.01in}R}(\text{i}\partial_{0}\overset{t_{x}}{\triangleright
}(u_{\bar{R},L})_{\ominus_{L}p,m}(x^{B},t_{x}))\nonumber\\
&  \quad=\,\kappa^{n}\int_{-\infty}^{+\infty}d_{1}^{n}p\,(u_{\bar{R}%
,L})_{\kappa p,m}(y^{A},-q^{\zeta}\kappa^{-2}t_{y})\overset{p|x}{\odot
}_{\hspace{-0.01in}R}(H_{0}\overset{x}{\triangleright}(u_{\bar{R},L}%
)_{\ominus_{L}p,m}(x^{B},t_{x}))\nonumber\\
&  \quad=\,H_{0}\overset{x}{\triangleright}(K_{1})_{m}^{\prime}(y^{i},x^{j}),
\label{GrenFrei1}%
\end{align}
which is a direct consequence of the fact that q-deformed plane waves are
solutions to Schr\"{o}dinger equations (for the details see part II). For the
last step in (\ref{SchrGlGre}) we need
\begin{align}
&  \lim_{t_{x}\rightarrow t_{y}}(K_{1})_{m}^{\prime}(y^{i},x^{j})\nonumber\\
&  \qquad=\,\kappa^{n}\int_{-\infty}^{+\infty}d_{1}^{n}p\,(u_{\bar{R}%
,L})_{\kappa p,m}(y^{A},-q^{\zeta}\kappa^{-2}t_{x})\overset{p|x}{\odot
}_{\hspace{-0.01in}R}(u_{\bar{R},L})_{\ominus_{L}p,m}(x^{B},t_{x})\nonumber\\
&  \qquad=\,\kappa^{n}\int_{-\infty}^{+\infty}d_{1}^{n}p\,(u_{\bar{R}%
,L})_{\kappa p,m}(y^{A},0)\overset{p|x}{\odot}_{\hspace{-0.01in}R}(u_{\bar
{R},L})_{\ominus_{L}p,m}(x^{B},0)\nonumber\\
&  \qquad=\,\kappa^{n}(\text{vol}_{1})^{-1}\,\delta_{1}^{n}(\kappa y^{A}%
\oplus_{\bar{R}}(\ominus_{\bar{R}}\,x^{B})). \label{BouConPro}%
\end{align}
The second step in (\ref{BouConPro}) uses that the time-dependent phase
factors of the plane waves cancel out against each other. Finally, the last
equality can be recognized as defining expression for a q-deformed delta
function as it was given in Ref. \cite{Qkin1}.

In the case of the advanced Green's functions similar arguments lead us to
\begin{align}
\text{i}\partial_{0}\overset{t_{x}}{\triangleright}(K_{1})_{m^{-}}^{\prime
}(y^{i},x^{j})=  &  \,-\,\text{i}\kappa^{n}(\text{vol}_{1})^{-1}\delta
(t_{y}-t_{x})\,\delta_{1}^{n}(\kappa y^{A}\oplus_{\bar{R}}(\ominus_{\bar{R}%
}\,x^{B}))\nonumber\\
&  \,-\,H_{0}\overset{x}{\triangleright}(K_{1})_{m^{-}}^{\prime}(y^{i},x^{j}).
\label{SchGleAvaGre}%
\end{align}
Repeating the same steps as above for the other geometries we additionally
obtain%
\begin{align}
&  \hspace{-0.4in}(K_{2})_{m^{\pm}}(x^{i},y^{j})\overset{t_{x}}{\triangleleft
}(\text{i}\hat{\partial}_{0})\nonumber\\
=  &  \mp\,\text{i}\kappa^{-n}(\text{vol}_{2})^{-1}\delta(\pm t_{x}\mp
t_{y})\,\delta_{2}^{n}((\ominus_{\bar{L}}\,x^{A})\oplus_{\bar{L}}(\kappa
^{-1}y^{B}))\nonumber\\
&  \mp\,(K_{2})_{m^{\pm}}(x^{i},y^{j})\overset{x}{\triangleleft}%
H_{0},\label{GrenFrei0}\\[0.08in]
&  \hspace{-0.4in}-\,\text{i}\partial_{0}\overset{t_{x}}{\triangleright}%
(K_{1}^{\ast})_{m^{\pm}}(x^{i},y^{j})\nonumber\\
=  &  \pm\,\text{i}\kappa^{n}(\text{vol}_{1})^{-1}\delta(\pm t_{x}\mp
t_{y})\,\delta_{1}^{n}(x^{A}\oplus_{\bar{R}}(\ominus_{\bar{R}}\,y^{B}%
))\nonumber\\
&  \pm\,H_{0}\overset{y}{\triangleright}(K_{1}^{\ast})_{m^{\pm}}(x^{i}%
,y^{j}),\\[0.08in]
&  \hspace{-0.4in}(K_{2}^{\ast})_{m^{\pm}}^{\prime}(y^{i},x^{j})\overset
{t_{x}}{\triangleleft}(\text{i}\hat{\partial}_{0})\nonumber\\
=  &  \mp\,\text{i}\kappa^{-n}(\text{vol}_{2})^{-1}\delta(\pm t_{x}\mp
t_{y})\,\delta_{2}^{n}((\ominus_{\bar{L}}\,y^{A})\oplus_{\bar{L}}%
x^{B})\nonumber\\
&  \mp\,(K_{2}^{\ast})_{m^{\pm}}(y^{i},x^{j})\overset{y}{\triangleleft}H_{0}.
\label{GrenFrei2}%
\end{align}

From the identities in (\ref{SchrGlGre}) and (\ref{GrenFrei0}) -
(\ref{GrenFrei2}) we can show that the Green's functions generate solutions to
the inhomogeneous Schr\"{o}dinger equations. In this manner, we have%
\begin{align}
\text{i}\partial_{0}\overset{t_{x}}{\triangleright}(\psi_{1})_{\varrho^{\pm}%
}^{\prime}(x^{i})\mp H_{0}\overset{x}{\triangleright}(\psi_{1})_{\varrho^{\pm
}}^{\prime}(x^{i})  &  =\varrho^{\pm}(x^{i}),\nonumber\\
\text{i}\partial_{0}\overset{t_{x}}{\triangleright}(\psi_{1}^{\ast}%
)_{\varrho^{\pm}}(x^{i})\mp H_{0}\overset{x}{\triangleright}(\psi_{1}^{\ast
})_{\varrho^{\pm}}(x^{i})  &  =\varrho^{\pm}(x^{i}),\\[0.16in]
(\psi_{2})_{\varrho^{\pm}}(x^{i})\overset{t_{x}}{\triangleleft}(\text{i}%
\hat{\partial}_{0})\pm(\psi_{2})_{\varrho^{\pm}}(x^{i})\overset{x}%
{\triangleleft}H_{0}  &  =\varrho^{\pm}(x^{i}),\nonumber\\
(\psi_{2}^{\ast})_{\varrho^{\pm}}^{\prime}(x^{i})\overset{t_{x}}%
{\triangleleft}(\text{i}\hat{\partial}_{0})\pm(\psi_{2}^{\ast})_{\varrho^{\pm
}}^{\prime}(x^{i})\overset{x}{\triangleleft}H_{0}  &  =\varrho^{\pm}(x^{i}),
\end{align}
with%
\begin{align}
(\psi_{1})_{\varrho^{\pm}}^{\prime}(x^{i})  &  =\mp\text{i}\int_{-\infty
}^{+\infty}dt_{y}\int_{-\infty}^{+\infty}d_{1}^{n}y\,\varrho^{\pm}%
(y^{j})\overset{y}{\circledast}(K_{1})_{m^{\pm}}^{\prime}(y^{k},x^{i}%
),\nonumber\\
(\psi_{2})_{\varrho^{\pm}}(x^{i})  &  =\pm\text{i}\int_{-\infty}^{+\infty
}dt_{y}\int\nolimits_{-\infty}^{+\infty}d_{2}^{n}y\,(K_{2})_{m^{\pm}}%
(x^{i},y^{j})\overset{y}{\circledast}\varrho^{\pm}(y^{k}), \label{SolInhSchr1}%
\\[0.1in]
(\psi_{1}^{\ast})_{\varrho^{\pm}}(x^{i})  &  =\mp\text{i}\int_{-\infty
}^{+\infty}dt_{y}\int\nolimits_{-\infty}^{+\infty}d_{1}^{n}y\,(K_{1}^{\ast
})_{m^{\pm}}(x^{i},y^{j})\overset{y}{\circledast}\varrho^{\pm}(\kappa
y^{A},t_{y}),\nonumber\\
(\psi_{2}^{\ast})_{\varrho^{\pm}}^{\prime}(x^{i})  &  =\pm\text{i}%
\int_{-\infty}^{+\infty}dt_{y}\int\nolimits_{-\infty}^{+\infty}d_{2}%
^{n}y\,\varrho^{\pm}(\kappa^{-1}y^{A},t_{y})\overset{y}{\circledast}%
(K_{2}^{\ast})_{m^{\pm}}^{\prime}(y^{k},x^{i}). \label{SolInhSchr2}%
\end{align}

These assertions can be checked as follows:%
\begin{align}
&  \text{i}\partial_{0}\overset{t_{x}}{\triangleright}(\psi_{1})_{\varrho
^{\pm}}^{\prime}(x^{i})\mp H_{0}\overset{x}{\triangleright}(\psi_{1}%
)_{\varrho^{\pm}}^{\prime}(x^{i})\nonumber\\
&  \qquad=\mp\text{i}\int_{-\infty}^{+\infty}dt_{y}\int_{-\infty}^{+\infty
}d_{1}^{n}y\,\varrho^{\pm}(y^{j})\overset{y}{\circledast}\big (\text{i}%
\partial_{0}\overset{t_{x}}{\triangleright}\mp H_{0}\overset{x}{\triangleright
}\big )(K_{1})_{m^{\pm}}^{\prime}(y^{k},x^{i})\nonumber\\
&  \qquad=\frac{\kappa^{n}}{\text{vol}_{1}}\int_{-\infty}^{+\infty}%
dt_{y}\,\delta(\pm t_{x}\mp t_{y})\int_{-\infty}^{+\infty}d_{1}^{n}%
y\,\varrho^{\pm}(y^{j})\overset{y}{\circledast}\delta_{1}^{n}(\kappa
y^{A}\oplus_{\bar{R}}(\ominus_{\bar{R}}\,x^{B}))\nonumber\\
&  \qquad=\frac{1}{\text{vol}_{1}}\int_{-\infty}^{+\infty}d_{1}^{n}%
y\,\varrho^{\pm}(\kappa^{-1}y^{C},t_{x})\overset{y}{\circledast}\delta_{1}%
^{n}(y^{A}\oplus_{\bar{R}}(\ominus_{\bar{R}}\,x^{B}))\nonumber\\
&  \qquad=\varrho^{\pm}(x^{i}).
\end{align}
For the first step we insert the expressions in (\ref{SolInhSchr1}) and
(\ref{SolInhSchr2}). The second equality is a consequence of (\ref{SchrGlGre})
and (\ref{SchGleAvaGre}). The time integral vanishes due to the delta function
and the last step makes use of the fundamental relations for q-deformed delta
functions, which were derived in Ref. \cite{Qkin2}.

It is also worth recording here that the q-deformed Green's functions show a
composition property. In analogy to their undeformed counterparts the
q-deformed Green's functions satisfy%
\begin{align}
(K_{1})_{m^{\pm}}^{\prime}(y^{i},x^{j})  &  =\int\nolimits_{-\infty}^{+\infty
}d_{1}^{n}z\,(K_{1})_{m^{\pm}}^{\prime}(y^{i},z^{k})\overset{z}{\circledast
}(K_{1})_{m^{\pm}}^{\prime}(z^{l},x^{j}),\nonumber\\
(K_{2})_{m^{\pm}}(x^{i},y^{j})  &  =\int\nolimits_{-\infty}^{+\infty}d_{2}%
^{n}z\,(K_{2})_{m^{\pm}}(x^{i},z^{k})\overset{z}{\circledast}(K_{2})_{m^{\pm}%
}(z^{l},y^{j}),\label{ComFree1}\\[0.1in]
(K_{1}^{\ast})_{m^{\pm}}(x^{i},y^{j})  &  =\int\nolimits_{-\infty}^{+\infty
}d_{1}^{n}z\,(K_{1}^{\ast})_{m^{\pm}}(x^{i},z^{k})\overset{z}{\circledast
}(K_{1}^{\ast})_{m^{\pm}}(\kappa z^{A},t_{z};y^{j}),\nonumber\\
(K_{2}^{\ast})_{m^{\pm}}^{\prime}(y^{i},x^{j})  &  =\int\nolimits_{-\infty
}^{+\infty}d_{2}^{n}z\,(K_{2}^{\ast})_{m^{\pm}}^{\prime}(y^{i};\kappa
^{-1}z^{A},t_{z})\overset{z}{\circledast}(K_{2}^{\ast})_{m^{\pm}}^{\prime
}(z^{l},x^{j}). \label{ComFree2}%
\end{align}
These identities can be proved in a rather straightforward manner. The
following calculation shall serve as an example:%
\begin{align}
&  (\phi_{1})_{m}^{\prime}(x^{i})=\int\nolimits_{-\infty}^{+\infty}d_{1}%
^{n}z\,(\phi_{1})_{m}^{\prime}(z^{j})\overset{z}{\circledast}(K_{1})_{m^{+}%
}^{\prime}(z^{k},x^{i})\nonumber\\
&  \qquad=\int\nolimits_{-\infty}^{+\infty}d_{1}^{n}z\,\int\nolimits_{-\infty
}^{+\infty}d_{1}^{n}y\,(\phi_{1})_{m}^{\prime}(y^{k})\overset{y}{\circledast
}(K_{1})_{m^{+}}^{\prime}(y^{l},z^{j})\overset{z}{\circledast}(K_{1})_{m^{+}%
}^{\prime}(z^{k},x^{i})\nonumber\\
&  \qquad=\int\nolimits_{-\infty}^{+\infty}d_{1}^{n}y\,(\phi_{1})_{m}^{\prime
}(y^{k})\overset{y}{\circledast}\int\nolimits_{-\infty}^{+\infty}d_{1}%
^{n}z\,(K_{1})_{m^{+}}^{\prime}(y^{l},z^{j})\overset{z}{\circledast}%
(K_{1})_{m^{+}}^{\prime}(z^{k},x^{i}).
\end{align}

Last but not least let us say a few words about the conjugation properties of
the Green's functions for the free non-relativistic particle on q-deformed
quantum spaces. Concretely, we have $(i=1,2)$%
\begin{align}
\overline{(K_{i})_{m^{\pm}}(x^{k},y^{l})}  &  =\theta(\pm t_{x}\mp
t_{y})\,(K_{i})_{m}^{\prime}(y^{B},\mp t_{y};x^{A},\mp t_{x})\equiv(\tilde
{K}_{i})_{m^{\pm}}^{\prime}(x^{k},y^{l}),\nonumber\\
\overline{(K_{i})_{m^{\pm}}^{\prime}(y^{k},x^{l})}  &  =\theta(\pm t_{x}\mp
t_{y})\,(K_{i})_{m}(x^{B},\pm t_{x};y^{A},\pm t_{y})\equiv(\tilde{K}%
_{i})_{m^{\pm}}(y^{k},x^{l}),\label{ConProp1}\\[0.1in]
\overline{(K_{i}^{\ast})_{m^{\pm}}(x^{k},y^{l})}  &  =\theta(\pm t_{x}\mp
t_{y})\,(K_{i}^{\ast})_{m}^{\prime}(y^{B},\pm t_{y};x^{A},\pm t_{x}%
)\equiv(\tilde{K}_{i}^{\ast})_{m^{\pm}}^{\prime}(x^{k},y^{l}),\nonumber\\
\overline{(K_{i}^{\ast})_{m^{\pm}}^{\prime}(y^{k},x^{l})}  &  =\theta(\pm
t_{x}\mp t_{y})\,(K_{i}^{\ast})_{m}(x^{B},\mp t_{x};y^{A},\mp t_{y}%
)\equiv(\tilde{K}_{i}^{\ast})_{m^{\pm}}(y^{k},x^{l}). \label{ConProp2}%
\end{align}
To check these relations, one can proceed as follows:%
\begin{align}
&  \overline{(K_{1})_{m^{+}}^{\prime}(y^{i},x^{j})}=\overline{\theta
(t_{x}-t_{y})(K_{1})_{m}^{\prime}(y^{i},x^{j})}=\theta(t_{x}-t_{y}%
)\overline{(K_{1})_{m}^{\prime}(y^{i},x^{j})}\nonumber\\
&  =\,\theta(t_{x}-t_{y})\kappa^{n}\!\int_{-\infty}^{+\infty}d_{1}%
^{n}p\,\overline{(u_{\bar{R},L})_{\kappa p,m}(y^{A},-q^{\zeta}\kappa^{-2}%
t_{y})\overset{p|x}{\odot}_{\hspace{-0.01in}R}(u_{\bar{R},L})_{\ominus_{L}%
p,m}(x^{B},t_{x})}\nonumber\\
&  =\,\theta(t_{x}-t_{y})\kappa^{n}\!\int_{-\infty}^{+\infty}d_{1}^{n}%
p\,(\bar{u}_{\bar{R},L})_{\ominus_{\bar{R}}p,m}(x^{B},t_{x})\overset
{p|x}{\odot}_{\hspace{-0.01in}\bar{L}}(\bar{u}_{\bar{R},L})_{\kappa p,m}%
(y^{A},-q^{\zeta}\kappa^{-2}t_{y})\nonumber\\
&  =\,\theta(t_{x}-t_{y})(K_{1})_{m}(x^{B},y^{A}).
\end{align}
Notice that for the third and fourth step we took into account the conjugation
properties of the elements of q-analysis, as they were given in Ref.
\cite{qAn}.

In (\ref{ConProp1}) and (\ref{ConProp2}) we introduced propagators with a
tilde. They belong to the instruction that particles with positive energy
travel backwards in time, while those with negative energy move forward in
time. In this sense, the new propagators do not conform with the usual
agreement in physics that particles with negative energy travel backwards in
time. However, this requirement is fulfilled\ by the propagators without a
tilde. In our formalism we decided to assign $(\phi_{i})_{m}^{\prime}(t)$ and
$(\phi_{i}^{\ast})_{m}(t)$ a positive energy, whereas $(\phi_{i}^{\ast}%
)_{m}^{\prime}(t)$ and $(\phi_{i})_{m}(t)$ correspond to negative energies.
Notice that in complete analogy to the undeformed case changing the sign of
the time variable from plus to minus transforms wave functions with positive
energy to those with negative energy and vice versa.

\section{Elements of scattering theory \label{ScaThe}}

\subsection{The Lippmann-Schwinger equations}

In scattering theory one typically considers the situation that the time
translation generator $H\ $can be divided into a free-particle Hamiltonian
$H_{0}$ and an interaction $V.$ However, in our formalism things become
slightly more difficult, since we have to distinguish the following versions
of time translation operators:%
\begin{equation}
H=H_{0}+V,\qquad H^{\prime}=q^{-\zeta}H_{0}+V(x^{A}),\qquad H^{\prime\prime
}=q^{\zeta}H_{0}+V(x^{A}).
\end{equation}
In this respect, it is our aim to seek solutions to the Schr\"{o}dinger
equations
\begin{align}
\text{i}\partial_{0}\overset{t}{\triangleright}(\psi_{1})_{m^{+}}^{\prime
}(x^{i})  &  =H^{\prime}\overset{x}{\triangleright}(\psi_{1})_{m^{+}}^{\prime
}(x^{i}),\nonumber\\
\text{i}\partial_{0}\overset{t}{\triangleright}(\psi_{1}^{\ast})_{m^{+}}%
(x^{i})  &  =H\overset{x}{\triangleright}(\psi_{1}^{\ast})_{m^{+}}%
(x^{i}),\label{SchrEqAlg1N}\\[0.1in]
\text{i}\hat{\partial}_{0}\,\overset{t}{\bar{\triangleright}}\,(\psi
_{2})_{m^{+}}^{\prime}(x^{i})  &  =H^{\prime\prime}\,\overset{x}%
{\bar{\triangleright}}\,(\psi_{2})_{m^{+}}^{\prime}(x^{i}),\nonumber\\
\text{i}\hat{\partial}_{0}\,\overset{t}{\bar{\triangleright}}\,(\psi_{2}%
^{\ast})_{m^{+}}(x^{i})  &  =H\,\overset{x}{\bar{\triangleright}}\,(\psi
_{2}^{\ast})_{m^{+}}(x^{i}),
\end{align}
and
\begin{align}
(\psi_{1})_{m^{-}}(x^{i})\,\overset{t}{\bar{\triangleleft}}\,(\text{i}%
\partial_{0})  &  =(\psi_{1})_{m^{-}}(x^{i})\,\overset{x}{\bar{\triangleleft}%
}\,H^{\prime},\nonumber\\
(\psi_{1}^{\ast})_{m^{-}}^{\prime}(x^{i})\,\overset{t}{\bar{\triangleleft}%
}\,(\text{i}\partial_{0})  &  =(\psi_{1}^{\ast})_{m^{-}}^{\prime}%
(x^{i})\,\overset{x}{\bar{\triangleleft}}\,H,\\[0.1in]
(\psi_{2})_{m^{-}}(x^{i})\overset{t}{\triangleleft}(\text{i}\hat{\partial}%
_{0})  &  =(\psi_{2})_{m^{-}}(x^{i})\overset{x}{\triangleleft}H^{\prime\prime
},\nonumber\\
(\psi_{2}^{\ast})_{m^{-}}^{\prime}(x^{i})\overset{t}{\triangleleft}%
(\text{i}\hat{\partial}_{0})  &  =(\psi_{2}^{\ast})_{m^{-}}^{\prime}%
(x^{i})\overset{x}{\triangleleft}H, \label{SchrEqAlg4N}%
\end{align}
where we require for\ these solutions to be subject to the boundary conditions
($i=1,2$)%
\begin{align}
\lim_{t_{x}\rightarrow+\infty}((\psi_{i})_{m}(x^{A},t_{x})-(\phi_{i}%
)_{m}(x^{A},t_{x}))  &  =0,\nonumber\\
\lim_{t_{x}\rightarrow-\infty}((\psi_{i})_{m}^{\prime}(x^{A},t_{x})-(\phi
_{i})_{m}^{\prime}(x^{A},t_{x}))  &  =0,\label{ConBoun1}\\[0.1in]
\lim_{t_{x}\rightarrow-\infty}((\psi_{i}^{\ast})_{m}(x^{A},t_{x})-(\phi
_{i}^{\ast})_{m}(x^{A},t_{x}))  &  =0,\nonumber\\
\lim_{t_{x}\rightarrow+\infty}((\psi_{i}^{\ast})_{m}^{\prime}(x^{A}%
,t_{x})-(\phi_{i}^{\ast})_{m}^{\prime}(x^{A},t_{x}))  &  =0. \label{ConBoun2}%
\end{align}

Essentially for us is the fact that the Schr\"{o}dinger equations in
(\ref{SchrEqAlg1N})-(\ref{SchrEqAlg4N}) can be seen as free-particle
Schr\"{o}dinger equations with an inhomogeneous contribution. In this manner
we have, for example,%
\begin{equation}
\text{i}\partial_{0}\overset{t_{x}}{\triangleright}(\psi_{1})_{m}^{\prime
}(x^{i})-q^{-\zeta}H_{0}\overset{x}{\triangleright}(\psi_{1})_{m}^{\prime
}(x^{i})=\varrho(x^{i}), \label{ExaInhSchr}%
\end{equation}
with%
\begin{equation}
\varrho(x^{i})=V(x^{j})\overset{x}{\circledast}(\psi_{1})_{m}^{\prime}%
(x^{i})=(\psi_{1})_{m}^{\prime}(x^{i})\overset{x}{\circledast}V(x^{j}).
\end{equation}
Recalling the identities in (\ref{SolInhSchr1}) and (\ref{SolInhSchr2}) we
find\ that a solution to (\ref{ExaInhSchr}) has to fulfill the so-called
Lippmann-Schwinger equation%
\begin{align}
&  (\psi_{1})_{m^{+}}^{\prime}(x^{i})=\,(\phi_{1})_{m}^{\prime}(x^{A}%
,q^{-\zeta}t_{x})\nonumber\\
&  -\text{i}\int_{-\infty}^{+\infty}dt_{y}\int_{-\infty}^{+\infty}d_{1}%
^{n}y\,(\psi_{1})_{m^{+}}^{\prime}(y^{j})\overset{y}{\circledast}%
V(y^{k})\overset{y}{\circledast}(K_{1})_{m^{+}}^{\prime}(y^{l};x^{A}%
,q^{-\zeta}t_{x}).
\end{align}

Notice that each solution to this integral equation shows the correct boundary
condition, since the Green's function vanishes as $t_{x}\rightarrow-\infty.$
Repeating the above arguments for the other geometries we also find%
\begin{align}
(\psi_{2})_{m^{-}}(x^{i})=\,  &  (\phi_{2})_{m}(x^{A},q^{\zeta}t_{x}%
)\nonumber\\
&  -\text{i}\int_{-\infty}^{+\infty}dt_{y}\int\nolimits_{-\infty}^{+\infty
}d_{2}^{n}y\,(K_{2})_{m^{-}}(x^{A},q^{\zeta}t_{x};y^{j})\nonumber\\
&  \qquad\qquad\qquad\qquad\overset{y}{\circledast}V(y^{k})\overset
{y}{\circledast}(\psi_{2})_{m^{-}}(y^{l}),\\[0.08in]
(\psi_{1}^{\ast})_{m^{+}}(x^{i})=\,  &  (\phi_{1}^{\ast})_{m}(x^{i}%
)\nonumber\\
&  -\text{i}\int_{-\infty}^{+\infty}dt_{y}\int\nolimits_{-\infty}^{+\infty
}d_{1}^{n}y\,(K_{1}^{\ast})_{m^{+}}(x^{i},y^{j})\overset{y}{\circledast
}V(\kappa y^{A},t_{y})\nonumber\\
&  \qquad\qquad\qquad\qquad\overset{y}{\circledast}(\psi_{1}^{\ast})_{m^{+}%
}(\kappa y^{B},t_{y}),\\[0.08in]
(\psi_{2}^{\ast})_{m^{-}}^{\prime}(x^{i})=\,  &  (\phi_{2}^{\ast})_{m}%
^{\prime}(x^{i})\nonumber\\
&  -\text{i}\int_{-\infty}^{+\infty}dt_{y}\int\nolimits_{-\infty}^{+\infty
}d_{2}^{n}y\,(\psi_{2}^{\ast})_{m^{-}}^{\prime}(\kappa^{-1}y^{A}%
,t_{y})\overset{y}{\circledast}V(\kappa^{-1}y^{B},t_{y})\nonumber\\
&  \qquad\qquad\qquad\qquad\overset{y}{\circledast}(K_{2}^{\ast})_{m^{-}%
}^{\prime}(y^{k},x^{i}).
\end{align}

To solve the Lippmann-Schwinger equations it is sometimes convenient to
introduce new Green's functions, for which we require to hold%
\begin{align}
(\psi_{1})_{m^{+}}^{\prime}(x^{i})  &  =\lim_{t_{y}\rightarrow-\infty}%
\int\nolimits_{-\infty}^{+\infty}d_{1}^{n}y\,(\phi_{1})_{m}^{\prime}%
(y^{A},q^{-\zeta}t_{y})\overset{y}{\circledast}(G_{1})_{m^{+}}^{\prime}%
(y^{B},t_{y};x^{i}),\nonumber\\
(\psi_{2})_{m^{-}}(x^{i})  &  =\lim_{t_{y}\rightarrow+\infty}\int
\nolimits_{-\infty}^{+\infty}d_{2}^{n}y\,(G_{2})_{m^{-}}(x^{i};y^{A}%
,t_{y})\overset{y}{\circledast}(\phi_{2})_{m}(y^{B},q^{\zeta}t_{y}%
),\label{DefGre1}\\[0.1in]
(\psi_{1}^{\ast})_{m^{+}}(x^{i})  &  =\lim_{t_{y}\rightarrow-\infty}%
\int\nolimits_{-\infty}^{+\infty}d_{1}^{n}y\,(G_{1}^{\ast})_{m^{+}}%
(x^{i};y^{A},t_{y})\overset{y}{\circledast}(\phi_{1}^{\ast})_{m}(\kappa
y^{B},t_{y}),\nonumber\\
(\psi_{2}^{\ast})_{m^{-}}^{\prime}(x^{i})  &  =\lim_{t_{y}\rightarrow+\infty
}\int\nolimits_{-\infty}^{+\infty}d_{2}^{n}y\,(\phi_{2}^{\ast})_{m}^{\prime
}(\kappa^{-1}y^{A},t_{y})\overset{y}{\circledast}(G_{2}^{\ast})_{m^{-}%
}^{\prime}(y^{B},t_{y};x^{i}). \label{DefGre2}%
\end{align}
With these relations at hand we can rewrite the Lippmann-Schwinger equations
in a way that enables us to read off equations for the new Green's functions:%
\begin{align}
&  (\psi_{1})_{m^{+}}^{\prime}(x^{i})=\,(\phi_{1})_{m}^{\prime}(x^{A}%
,q^{-\zeta}t_{x})\nonumber\\
&  \quad\,\hspace{0.18in}-\,\text{i}\int_{-\infty}^{+\infty}dt_{y}%
\int_{-\infty}^{+\infty}d_{1}^{n}y\,(\psi_{1})_{m^{+}}^{\prime}(y^{j}%
)\overset{y}{\circledast}V(y^{k})\overset{y}{\circledast}(K_{1})_{m^{+}%
}^{\prime}(y^{l};x^{A},q^{-\zeta}t_{x})\nonumber\\
&  \quad\,=\,\lim_{t_{y_{1}}\rightarrow-\infty}\int\nolimits_{-\infty
}^{+\infty}d_{1}^{n}y_{1}\,(\phi_{1})_{m}^{\prime}(y_{1}^{A},t_{y}%
)\overset{y_{1}}{\circledast}(K_{1})_{m^{+}}^{\prime}(y_{1}^{B},t_{y_{1}%
};x^{C},q^{-\zeta}t_{x})\nonumber\\
&  \quad\,\hspace{0.18in}-\,\lim_{t_{y_{1}}\rightarrow-\infty}\text{i}%
\int_{-\infty}^{+\infty}dt_{y_{2}}\int_{-\infty}^{+\infty}d_{1}^{n}y_{2}%
\,\int\nolimits_{-\infty}^{+\infty}d_{1}^{n}y_{1}\,(\phi_{1})_{m}^{\prime
}(y_{1}^{A},t_{y_{1}})\nonumber\\
&  \quad\,\hspace{0.18in}\qquad\qquad\hspace{0.18in}\overset{y_{1}%
}{\circledast}(G_{1})_{m^{+}}^{\prime}(y_{1}^{B},t_{y_{1}};y_{2}^{j}%
)\overset{y_{2}}{\circledast}V(y_{2}^{k})\overset{y_{2}}{\circledast}%
(K_{1})_{m^{+}}^{\prime}(y_{2}^{l};x^{C},q^{-\zeta}t_{x}).
\end{align}
Comparing this result with the first relation in (\ref{DefGre1}) finally
yields%
\begin{align}
&  (G_{1})_{m^{+}}^{\prime}(y^{i},x^{j})=\,(K_{1})_{m^{+}}^{\prime}%
(y^{i};x^{A},q^{-\zeta}t_{x})\nonumber\\
&  \quad-\,\text{i}\int_{-\infty}^{+\infty}dt_{z}\int\nolimits_{-\infty
}^{+\infty}d_{1}^{n}z\,(G_{1})_{m^{+}}^{\prime}(y^{i},z^{k})\overset
{z}{\circledast}V(z^{l})\overset{z}{\circledast}(K_{1})_{m^{+}}^{\prime}%
(z^{r};x^{A},q^{-\zeta}t_{x}). \label{GrenWech1}%
\end{align}
Applying similar arguments to the other geometries leads us to
\begin{align}
&  (G_{2})_{m^{-}}(x^{i},y^{j})=\,(K_{2})_{m^{-}}(x^{A},q^{\zeta}t_{x}%
;y^{j})\nonumber\\
&  \qquad-\,\text{i}\int_{-\infty}^{+\infty}dt_{z}\int\nolimits_{-\infty
}^{+\infty}d_{2}^{n}z\,(K_{2})_{m^{-}}(x^{A},q^{\zeta}t_{x};,z^{k})\overset
{z}{\circledast}V(z^{l})\nonumber\\
&  \qquad\qquad\qquad\qquad\overset{z}{\circledast}(G_{2})_{m^{-}}(z^{r}%
,y^{j}),\\[0.1in]
&  (G_{1}^{\ast})_{m^{+}}(x^{i},y^{j})=\,(K_{1}^{\ast})_{m^{+}}(x^{i}%
,y^{j})\nonumber\\
&  \qquad-\,\text{i}\int_{-\infty}^{+\infty}dt_{z}\int\nolimits_{-\infty
}^{+\infty}d_{1}^{n}z\,(K_{1}^{\ast})_{m^{+}}(x^{i},z^{k})\overset
{z}{\circledast}V(\kappa z^{A},t_{z})\nonumber\\
&  \qquad\qquad\qquad\qquad\overset{z}{\circledast}(G_{1}^{\ast})_{m^{+}%
}(\kappa z^{B},t_{z};y^{j}),\\[0.1in]
&  (G_{2}^{\ast})_{m^{-}}^{\prime}(y^{i},x^{j})=\,(K_{2}^{\ast})_{m^{-}%
}^{\prime}(y^{i},x^{j})\nonumber\\
&  \qquad-\,\text{i}\int_{-\infty}^{+\infty}dt_{z}\int\nolimits_{-\infty
}^{+\infty}d_{2}^{n}z\,(G_{2}^{\ast})_{m^{-}}^{\prime}(y^{i},\kappa^{-1}%
z^{A},t_{z})\overset{z}{\circledast}V(\kappa^{-1}z^{B},t_{z})\nonumber\\
&  \qquad\qquad\qquad\qquad\overset{z}{\circledast}(K_{2}^{\ast})_{m^{-}%
}^{\prime}(z^{k},x^{j}). \label{GrenWech2}%
\end{align}

Next, we would like to mention that the Green's functions satisfy%
\begin{align}
&  \text{i}\partial_{0}\overset{t_{x}}{\triangleright}(G_{1})_{m^{+}}^{\prime
}(y^{i},x^{j})-H^{\prime}\overset{x}{\triangleright}(G_{1})_{m^{+}}^{\prime
}(y^{i},x^{j})=\nonumber\\
&  \qquad\qquad=\,\text{i}q^{-\zeta}\kappa^{n}(\text{vol}_{1})^{-1}%
\delta(q^{-\zeta}t_{x}-t_{y})\,\delta_{1}^{n}(\kappa y^{A}\oplus_{\bar{R}%
}(\ominus_{\bar{R}}\,x^{B})),\label{ChaProGre1}\\[0.08in]
&  (G_{2})_{m^{-}}(x^{i},y^{j})\overset{t_{x}}{\triangleleft}(\text{i}%
\hat{\partial}_{0})-(G_{2})_{m^{-}}(x^{i},y^{j})\overset{x}{\triangleleft
}H^{\prime\prime}=\nonumber\\
&  \qquad\qquad=\,\text{i}q^{\zeta}\kappa^{-n}(\text{vol}_{2})^{-1}%
\delta(t_{y}-q^{\zeta}t_{x})\,\delta_{2}^{n}((\ominus_{\bar{L}}\,x^{A}%
)\oplus_{\bar{L}}(\kappa^{-1}y^{B})),\\[0.08in]
&  \text{i}\partial_{0}\overset{t_{y}}{\triangleright}(G_{1}^{\ast})_{m^{+}%
}(x^{i},y^{j})-H\overset{y}{\triangleright}(G_{1}^{\ast})_{m^{+}}(x^{i}%
,y^{j})=\nonumber\\
&  \qquad\qquad=\,\text{i}\kappa^{n}(\text{vol}_{1})^{-1}\delta(t_{x}%
-t_{y})\,\delta_{1}^{n}(x^{A}\oplus_{\bar{R}}(\ominus_{\bar{R}}\,y^{B}%
)),\\[0.08in]
&  (G_{2}^{\ast})_{m^{-}}^{\prime}(y^{i},x^{j})\overset{t_{y}}{\triangleleft
}(\text{i}\hat{\partial}_{0})-(G_{2}^{\ast})_{m^{-}}^{\prime}(y^{i}%
,x^{j})\overset{y}{\triangleleft}H=\nonumber\\
&  \qquad\qquad=\,\text{i}\kappa^{-n}(\text{vol}_{2})^{-1}\delta(t_{y}%
-t_{x})\,\delta_{2}^{n}((\ominus_{\bar{L}}\,y^{A})\oplus_{\bar{L}}x^{B}).
\end{align}
To prove these identities we first substitute the expressions in
(\ref{GrenWech1})-(\ref{GrenWech2}) for the Green's functions and then apply
the relations in (\ref{SchrGlGre}) and (\ref{GrenFrei0})-(\ref{GrenFrei2}).

It is also worth recording here that the Green's functions defined by the
relations in (\ref{DefGre1}) and (\ref{DefGre2}) can alternatively be
introduced by%
\begin{align}
&  q^{-\zeta}\theta(q^{-\zeta}t_{x}-t_{y})(\psi_{1})_{m^{+}}^{\prime}%
(x^{i})=\nonumber\\
&  \qquad\qquad=\int\nolimits_{-\infty}^{+\infty}d_{1}^{n}y\,(\psi_{1}%
)_{m^{+}}^{\prime}(y^{A},t_{y})\overset{y}{\circledast}(G_{m^{+}})_{1}%
(y^{B},t_{y};x^{i}),\label{AltDefGree0}\\[0.08in]
&  q^{\zeta}\theta(t_{y}-q^{\zeta}t_{x})(\psi_{2})_{m^{-}}(x^{i})=\nonumber\\
&  \qquad\qquad=\int\nolimits_{-\infty}^{+\infty}d_{2}^{n}y\,(G_{m^{-}}%
)_{2}(x^{i};y^{A},t_{y})\overset{y}{\circledast}(\psi_{2})_{m^{-}}(y^{B}%
,t_{y}),\label{AltDefGree1}\\[0.08in]
&  \theta(t_{x}-t_{y})(\psi_{1}^{\ast})_{m^{+}}(x^{i})=\nonumber\\
&  \qquad\qquad=\int\nolimits_{-\infty}^{+\infty}d_{1}^{n}y\,(G_{m^{+}}^{\ast
})_{1}(x^{i};y^{A},t_{y})\overset{y}{\circledast}(\psi_{1}^{\ast})_{m^{+}%
}(\kappa y^{B},t_{y}),\\[0.08in]
&  \theta(t_{y}-t_{x})(\psi_{2}^{\ast})_{m^{-}}^{\prime}(x^{i})=\nonumber\\
&  \qquad\qquad=\int\nolimits_{-\infty}^{+\infty}d_{2}^{n}y\,(\psi_{2}^{\ast
})_{m^{-}}^{\prime}(\kappa^{-1}y^{A},t_{y})\overset{y}{\circledast}(G_{m^{-}%
}^{\ast})_{2}^{\prime}(y^{B},t_{y};x^{i}). \label{AltDefGree2}%
\end{align}

To show that these definitions are indeed equivalent to those in
(\ref{DefGre1}) and (\ref{DefGre2}) we apply Schr\"{o}dinger operators to both
sides of each equality in (\ref{AltDefGree0}) and (\ref{AltDefGree2}). This
way, we obtain, for example,\
\begin{align}
&  \big (\text{i}\partial_{0}\overset{t_{x}}{\triangleright}-H^{\prime
}\overset{x}{\triangleright}\big )\big (\theta(q^{-\zeta}t_{x}-t_{y})(\psi
_{1})_{m^{+}}^{\prime}(x^{i})\big )=\nonumber\\
&  \qquad=\,\text{i}q^{-\zeta}\delta(q^{-\zeta}t_{x}-t_{y})(\psi_{1})_{m^{+}%
}^{\prime}(x^{i})+\theta(t_{x}-t_{y})\big (\text{i}\partial_{0}\overset{t_{x}%
}{\triangleright}-H\overset{x}{\triangleright}\big )(\psi_{1})_{m^{+}}%
^{\prime}(x^{i})\nonumber\\
&  \qquad=\,\text{i}q^{-\zeta}\delta(q^{-\zeta}t_{x}-t_{y})(\psi_{1})_{m^{+}%
}^{\prime}(x^{A},t_{x}).
\end{align}
Due to the characteristic property of the Green's function [cf. Eq.
(\ref{ChaProGre1})] this is equal to the expression%
\begin{align}
&  \big (\text{i}\partial_{0}\overset{t_{x}}{\triangleright}-H^{\prime
}\overset{x}{\triangleright}\big )\int\nolimits_{-\infty}^{+\infty}d_{1}%
^{n}y\,(\psi_{1})_{m^{+}}^{\prime}(y^{A},t_{y})\overset{y}{\circledast}%
(G_{1})_{m^{+}}^{\prime}(y^{B},t_{y};x^{i})=\nonumber\\
&  \quad=\,\int\nolimits_{-\infty}^{+\infty}d_{1}^{n}y\,(\psi_{1})_{m}%
^{\prime}(y^{A},t_{y})\overset{y}{\circledast}\big (\text{i}\partial
_{0}\overset{t_{x}}{\triangleright}-H^{\prime}\overset{x}{\triangleright
}\big )(G_{1})_{m^{+}}^{\prime}(y^{B},t_{y};x^{i})\nonumber\\
&  \quad=\,\text{i}\delta(q^{-\zeta}t_{x}-t_{y})\,\frac{q^{-\zeta}\kappa^{n}%
}{\text{vol}_{1}}\int\nolimits_{-\infty}^{+\infty}d_{1}^{n}y\,(\psi
_{1})_{m^{+}}^{\prime}(y^{A},t_{y})\overset{y}{\circledast}\delta_{1}%
^{n}(\kappa y^{B}\oplus_{\bar{R}}(\ominus_{\bar{R}}\,x^{C}))\nonumber\\
&  \quad=\,\text{i}q^{-\zeta}\delta(q^{-\zeta}t_{x}-t_{y})\,(\psi_{1})_{m^{+}%
}^{\prime}(x^{A},t_{x}).
\end{align}

The Green's functions defined by the relations in (\ref{DefGre1}) and
(\ref{DefGre2}) again show the composition property. Concretely, it holds%
\begin{align}
&  (G_{1})_{m^{+}}^{\prime}(y^{A},t_{y};x^{B},t_{x})=\nonumber\\
&  \quad=\int\nolimits_{-\infty}^{+\infty}d_{1}^{n}z\,(G_{1})_{m^{+}}^{\prime
}(y^{A},t_{y};z^{C},t_{z})\overset{z}{\circledast}(G_{1})_{m^{+}}^{\prime
}(z^{D},t_{z};x^{B},t_{x}),\label{Comp1}\\[0.1in]
&  (G_{2})_{m^{-}}(y^{A},t_{y};x^{B},t_{x})=\nonumber\\
&  \quad=\int\nolimits_{-\infty}^{+\infty}d_{2}^{n}z\,(G_{2})_{m^{-}}%
(y^{A},t_{y};z^{C},t_{z})\overset{z}{\circledast}(G_{2})_{m^{-}}(z^{D}%
,t_{z};x^{B},t_{x}),
\end{align}
and%
\begin{align}
&  (G_{1}^{\ast})_{m^{+}}(x^{A},t_{x};y^{B},t_{y})=\nonumber\\
&  \quad=\int\nolimits_{-\infty}^{+\infty}d_{1}^{n}z\,(G_{1}^{\ast})_{m^{+}%
}(x^{A},t_{x};z^{C},t_{z})\overset{z}{\circledast}(G_{1}^{\ast})_{m^{+}%
}(\kappa z^{D},t_{z};y^{B},t_{y}),\\[0.1in]
&  (G_{2}^{\ast})_{m^{-}}^{\prime}(x^{A},t_{x};y^{B},t_{y})=\nonumber\\
&  \quad=\int\nolimits_{-\infty}^{+\infty}d_{2}^{n}z\,(G_{2}^{\ast})_{m^{-}%
}^{\prime}(x^{A},t_{x};\kappa^{-1}z^{C},t_{z})\overset{z}{\circledast}%
(G_{2}^{\ast})_{m^{-}}^{\prime}(z^{D},t_{z};y^{B},t_{y}), \label{Comp2N}%
\end{align}
where $t_{x}>t_{z}>t_{y}.$ In addition to this, we have the identities%
\begin{align}
&  \int\nolimits_{-\infty}^{+\infty}d_{1}^{n}z\,(\tilde{G}_{1})_{m^{-}%
}^{\prime}(y^{A},t;z^{C},t_{-})\overset{z}{\circledast}(G_{1})_{m^{+}}%
^{\prime}(z^{D},t_{-};x^{B},t)=\nonumber\\
&  \qquad=\,\int\nolimits_{-\infty}^{+\infty}d_{1}^{n}z\,(G_{1})_{m^{+}%
}^{\prime}(y^{A},t;z^{C},t_{+})\overset{z}{\circledast}(\tilde{G}_{1})_{m^{-}%
}^{\prime}(z^{D},t_{+};x^{B},t)\nonumber\\[0.08in]
&  \qquad=\,\kappa^{n}(\text{vol}_{1})^{-1}\,\delta_{1}^{n}(\kappa y^{A}%
\oplus_{\bar{R}}(\ominus_{\bar{R}}\,x^{B})),\label{GreeInv1N}\\[0.1in]
&  \int\nolimits_{-\infty}^{+\infty}d_{2}^{n}z\,(\tilde{G}_{1})_{m^{+}}%
(y^{A},t;z^{C},t_{-})\overset{z}{\circledast}(G_{2})_{m^{-}}(z^{D},t_{-}%
;x^{B},t)=\nonumber\\
&  \qquad=\,\int\nolimits_{-\infty}^{+\infty}d_{2}^{n}z\,(G_{2})_{m^{-}}%
(y^{A},t;z^{C},t_{+})\overset{z}{\circledast}(\tilde{G}_{2})_{m^{+}}%
(z^{D},t_{+};x^{B},t)\nonumber\\
&  \qquad=\,\kappa^{-n}(\text{vol}_{2})^{-1}\,\delta_{2}^{n}((\ominus_{\bar
{L}}\,y^{A})\oplus_{\bar{L}}(\kappa^{-1}x^{B})), \label{GreeInv1}%
\end{align}
and%
\begin{align}
&  \int\nolimits_{-\infty}^{+\infty}d_{1}^{n}z\,(\tilde{G}_{1}^{\ast})_{m^{-}%
}(x^{A},t;z^{C},t_{+})\overset{z}{\circledast}(G_{1}^{\ast})_{m^{+}}(\kappa
z^{D},t_{+};y^{B},t)=\nonumber\\
&  \qquad=\,\int\nolimits_{-\infty}^{+\infty}d_{1}^{n}z\,(G_{1}^{\ast}%
)_{m^{+}}(x^{A},t;z^{C},t_{-})\overset{z}{\circledast}(\tilde{G}_{1}^{\ast
})_{m^{-}}(\kappa z^{D},t_{-};y^{B},t)\nonumber\\
&  \qquad=\,\kappa^{n}(\text{vol}_{1})^{-1}\,\delta_{1}^{n}(x^{A}\oplus
_{\bar{R}}(\ominus_{\bar{R}}\,y^{B})),\\[0.1in]
&  \int\nolimits_{-\infty}^{+\infty}d_{2}^{n}z\,(\tilde{G}_{2}^{\ast})_{m^{+}%
}^{\prime}(x^{A},t;\kappa^{-1}z^{C},t_{+})\overset{z}{\circledast}(G_{2}%
^{\ast})_{m^{-}}^{\prime}(z^{D},t_{+};y^{B},t)=\nonumber\\
&  \qquad=\,\int\nolimits_{-\infty}^{+\infty}d_{2}^{n}z\,(G_{2}^{\ast}%
)_{m^{-}}^{\prime}(x^{A},t;\kappa^{-1}z^{C},t_{-})\overset{z}{\circledast
}(\tilde{G}_{2}^{\ast})_{m^{+}}^{\prime}(z^{D},t_{-};y^{B},t)\nonumber\\
&  \qquad=\,\kappa^{-n}(\text{vol}_{2})^{-1}\,\delta_{2}^{n}((\ominus_{\bar
{L}}\,x^{A})\oplus_{\bar{L}}y^{B}), \label{GreeInv2N}%
\end{align}
where we now assume $t_{+}>t>t_{-}.$

Notice that the expressions for the Green's functions with a tilde are
obtained from those in (\ref{GrenWech1})-(\ref{GrenWech2}) by replacing the
free-particle Green's functions with the corresponding ones in (\ref{ConProp1}%
) and (\ref{ConProp2}):%
\begin{align}
&  (\tilde{G}_{1})_{m^{-}}^{\prime}(y^{i},x^{j})=\,(\tilde{K}_{1})_{m^{-}%
}^{\prime}(y^{i};x^{A},q^{-\zeta}t_{x})\nonumber\\
&  \qquad-\,\text{i}\int_{-\infty}^{+\infty}dt_{z}\int\nolimits_{-\infty
}^{+\infty}d_{1}^{n}z\,(\tilde{G}_{1})_{m^{+}}^{\prime}(y^{i},z^{k}%
)\overset{z}{\circledast}V(z^{l})\nonumber\\
&  \qquad\qquad\qquad\qquad\overset{z}{\circledast}(\tilde{K}_{1})_{m^{-}%
}^{\prime}(z^{r};x^{A},q^{-\zeta}t_{x}),\\[0.1in]
&  (\tilde{G}_{2})_{m^{+}}(x^{i},y^{j})=\,(\tilde{K}_{2})_{m^{+}}%
(x^{A},q^{\zeta}t_{x};y^{j})\nonumber\\
&  \qquad-\,\text{i}\int_{-\infty}^{+\infty}dt_{z}\int\nolimits_{-\infty
}^{+\infty}d_{2}^{n}z\,(\tilde{K}_{2})_{m^{+}}(x^{A},q^{\zeta}t_{x}%
;z^{k})\overset{z}{\circledast}V(z^{l})\nonumber\\
&  \qquad\qquad\qquad\qquad\overset{z}{\circledast}(\tilde{G}_{2})_{m^{+}%
}(z^{r},y^{j}),\\[0.1in]
&  (\tilde{G}_{1}^{\ast})_{m^{-}}(x^{i},y^{j})=\,(\tilde{K}_{1}^{\ast}%
)_{m^{-}}(x^{i},y^{j})\nonumber\\
&  \qquad-\,\text{i}\int_{-\infty}^{+\infty}dt_{z}\int\nolimits_{-\infty
}^{+\infty}d_{1}^{n}z\,(\tilde{K}_{1}^{\ast})_{m^{-}}(x^{i},z^{k})\overset
{z}{\circledast}V(\kappa z^{A},t_{z})\nonumber\\
&  \qquad\qquad\qquad\qquad\overset{z}{\circledast}(\tilde{G}_{1}^{\ast
})_{m^{-}}(\kappa z^{B},t_{z};y^{j}),\\[0.1in]
&  (\tilde{G}_{2}^{\ast})_{m^{+}}^{\prime}(y^{i},x^{j})=\,(\tilde{K}_{2}%
^{\ast})_{m^{+}}^{\prime}(y^{i},x^{j})\nonumber\\
&  \qquad-\,\text{i}\int_{-\infty}^{+\infty}dt_{z}\int\nolimits_{-\infty
}^{+\infty}d_{2}^{n}z\,(\tilde{G}_{2}^{\ast})_{m^{+}}^{\prime}(y^{i}%
,\kappa^{-1}z^{A},t_{z})\overset{z}{\circledast}V(\kappa^{-1}z^{B}%
,t_{z})\nonumber\\
&  \qquad\qquad\qquad\qquad\overset{z}{\circledast}(\tilde{K}_{2}^{\ast
})_{m^{+}}^{\prime}(z^{k},x^{j}).
\end{align}

The relations in (\ref{Comp1})-(\ref{Comp2N}) follow from the same reasonings
as those in (\ref{ComFree1})-(\ref{ComFree2}) if we take into account the
identities in (\ref{AltDefGree1}) and (\ref{AltDefGree2}) . To prove the
relations in (\ref{GreeInv1N})-(\ref{GreeInv2N}), we can proceed as follows:
\begin{align}
\psi(x^{i})  &  =\int\nolimits_{-\infty}^{+\infty}d_{1}^{n}y\,\psi(y^{C}%
,t_{+})\overset{y}{\circledast}(\tilde{G}_{1})_{m^{-}}^{\prime}(y^{B}%
,t_{+};x^{A},t_{x})\nonumber\\
&  =\int\nolimits_{-\infty}^{+\infty}d_{1}^{n}z\int\nolimits_{-\infty
}^{+\infty}d_{1}^{n}y\,\psi(z^{E},t)\overset{z}{\circledast}(G_{1})_{m^{+}%
}^{\prime}(z^{D},t_{x};y^{C},t_{+})\nonumber\\
&  \qquad\qquad\overset{y}{\circledast}(\tilde{G}_{1})_{m^{-}}^{\prime}%
(y^{B},t_{+};x^{A},t_{x})\nonumber\\
&  =\int\nolimits_{-\infty}^{+\infty}d_{1}^{n}z\,\psi(z^{E},t_{x})\overset
{z}{\circledast}\int\nolimits_{-\infty}^{+\infty}d_{1}^{n}y\,(G_{1})_{m^{+}%
}^{\prime}(z^{D},t_{x};y^{C},t_{+})\nonumber\\
&  \qquad\qquad\overset{y}{\circledast}(\tilde{G}_{1})_{m^{-}}^{\prime}%
(y^{B},t_{+};x^{A},t_{x}).
\end{align}
On the other hand, we have (see for example Ref. \cite{Qkin1})%
\begin{equation}
\psi(x^{i})=\frac{\kappa^{n}}{\text{vol}_{1}}\int\nolimits_{-\infty}^{+\infty
}d_{1}^{n}z\,\psi(z^{E},t_{x})\overset{z}{\circledast}\delta_{1}^{n}(\kappa
z^{D}\oplus_{\bar{R}}(\ominus_{\bar{R}}\,x^{A})),
\end{equation}
which shows us the validity of the first relation in (\ref{GreeInv1}).

Let us return to the Lippmann-Schwinger equations, once again. If the
interaction $V$ is small, we can solve them iteratively. In doing so we obtain
q-analogs of the famous Born series:%
\begin{align}
&  (\psi_{1})_{m^{+}}^{\prime}(x^{i})=\,(\phi_{1})_{m}^{\prime}(x^{A}%
,q^{-\zeta}t_{x})\nonumber\\
&  \quad+\,\text{i}^{-1}\int_{-\infty}^{+\infty}dt_{1}\int_{-\infty}^{+\infty
}d_{1}^{n}y_{1}\,(\phi_{1})_{m}^{\prime}(y_{1}^{A},q^{-\zeta}t_{1}%
)\overset{y_{1}}{\circledast}V(y_{1}^{j})\nonumber\\
&  \qquad\qquad\qquad\overset{y_{1}}{\circledast}(K_{1})_{m^{+}}^{\prime
}(y_{1}^{k};x^{B},q^{-\zeta}t_{x})\nonumber\\
&  \quad+\,\text{i}^{-2}\int_{-\infty}^{+\infty}dt_{1}\int_{-\infty}^{+\infty
}d_{1}^{n}y_{1}\,\int_{-\infty}^{+\infty}dt_{2}\int_{-\infty}^{+\infty}%
d_{1}^{n}y_{2}\,(\phi_{1})_{m}^{\prime}(y_{2}^{A},q^{-\zeta}t_{2})\nonumber\\
&  \qquad\qquad\qquad\overset{y_{2}}{\circledast}V(y_{2}^{j})\overset{y_{2}%
}{\circledast}(K_{1})_{m^{+}}^{\prime}(y_{2}^{k};y_{1}^{B},q^{-\zeta}%
t_{1})\overset{y_{1}}{\circledast}V(y_{1}^{l})\nonumber\\
&  \qquad\qquad\qquad\overset{y_{1}}{\circledast}(K_{1})_{m^{+}}^{\prime
}(y_{1}^{r};x^{C},q^{-\zeta}t_{x})+\ldots,\label{SolInt1}\\[0.1in]
&  (\psi_{2})_{m^{-}}(x^{i})=\,(\phi_{2})_{m}(x^{A},q^{\zeta}t_{x})\nonumber\\
&  \quad+\,\text{i}^{-1}\int_{-\infty}^{+\infty}dt_{1}\int\nolimits_{-\infty
}^{+\infty}d_{2}^{n}y_{1}\,(K_{2})_{m^{-}}(x^{A},q^{\zeta}t_{x};y_{1}%
^{j})\overset{y_{1}}{\circledast}V(y_{1}^{k})\nonumber\\
&  \qquad\qquad\qquad\overset{y_{1}}{\circledast}(\phi_{2})_{m}^{\prime}%
(y_{1}^{B},q^{\zeta}t_{1})\nonumber\\
&  \quad+\,\text{i}^{-2}\int_{-\infty}^{+\infty}dt_{1}\int\nolimits_{-\infty
}^{+\infty}d_{2}^{n}y_{1}\,\int_{-\infty}^{+\infty}dt_{2}\int
\nolimits_{-\infty}^{+\infty}d_{2}^{n}y_{2}\,(K_{2})_{m^{-}}(x^{A},q^{\zeta
}t_{x};y_{1}^{j})\nonumber\\
&  \qquad\qquad\qquad\overset{y_{1}}{\circledast}V(y_{1}^{k})\overset{y_{1}%
}{\circledast}(K_{2})_{m^{-}}(y_{1}^{B},q^{\zeta}t_{1};y_{2}^{l}%
)\overset{y_{2}}{\circledast}V(y_{2}^{r})\nonumber\\
&  \qquad\qquad\qquad\overset{y_{2}}{\circledast}(\phi_{2})_{m}(y_{2}%
^{C},q^{\zeta}t_{2})+\ldots,
\end{align}
and%
\begin{align}
&  (\psi_{1}^{\ast})_{m^{+}}(x^{i})=\,(\phi_{1}^{\ast})_{m}(x^{i})\nonumber\\
&  \hspace{0.05in}+\,\text{i}^{-1}\int_{-\infty}^{+\infty}dt_{1}\int_{-\infty
}^{+\infty}d_{1}^{n}y_{1}\,(K_{1}^{\ast})_{m^{+}}(x^{i},y_{1}^{j}%
)\overset{y_{1}}{\circledast}V(\kappa y_{1}^{A},t_{_{1}})\overset{y_{1}%
}{\circledast}(\phi_{1}^{\ast})_{m}(\kappa y_{1}^{B},t_{1})\nonumber\\
&  \hspace{0.05in}+\,\text{i}^{-2}\int_{-\infty}^{+\infty}dt_{1}\int_{-\infty
}^{+\infty}d_{1}^{n}y_{1}\,\int_{-\infty}^{+\infty}dt_{2}\int_{-\infty
}^{+\infty}d_{1}^{n}y_{2}\,(K_{1}^{\ast})_{m^{+}}(x^{i},y_{1}^{j})\nonumber\\
&  \hspace{0.05in}\qquad\qquad\quad\overset{y_{1}}{\circledast}V(\kappa
y_{1}^{A},t_{_{1}})\overset{y_{1}}{\circledast}(K_{1}^{\ast})_{m^{+}}(\kappa
y_{1}^{B},t_{1};y_{2}^{k})\overset{y_{2}}{\circledast}V(\kappa y_{2}^{C}%
,t_{2})\nonumber\\
&  \hspace{0.05in}\qquad\qquad\quad\overset{y_{2}}{\circledast}(\phi_{1}%
^{\ast})_{m}(\kappa y_{2}^{D},t_{2})+\ldots,\\[0.1in]
&  (\psi_{2}^{\ast})_{m}^{\prime}(x^{i})=\,(\phi_{2}^{\ast})_{m}^{\prime
}(x^{i})\nonumber\\
&  \hspace{0.05in}+\,\text{i}^{-1}\int_{-\infty}^{+\infty}dt_{1}%
\int\nolimits_{-\infty}^{+\infty}d_{2}^{n}y_{1}\,(\phi_{2}^{\ast})_{m}%
^{\prime}(\kappa^{-1}y_{1}^{A},t_{1})\overset{y_{1}}{\circledast}V(\kappa
^{-1}y_{1}^{B},t_{1})\nonumber\\
&  \qquad\qquad\quad\overset{y_{1}}{\circledast}(K_{2}^{\ast})_{m^{-}}%
^{\prime}(y_{1}^{k},x^{i})\nonumber\\
&  \hspace{0.05in}+\,\text{i}^{-2}\int_{-\infty}^{+\infty}dt_{1}%
\int\nolimits_{-\infty}^{+\infty}d_{2}^{n}y_{1}\,\int_{-\infty}^{+\infty
}dt_{2}\int\nolimits_{-\infty}^{+\infty}d_{2}^{n}y_{2}\,(\phi_{2}^{\ast}%
)_{m}^{\prime}(\kappa^{-1}y_{2}^{A},t_{2})\nonumber\\
&  \hspace{0.05in}\qquad\qquad\quad\overset{y_{2}}{\circledast}V(\kappa
y_{2}^{B},t_{2})\overset{y_{2}}{\circledast}(K_{2}^{\ast})_{m^{-}}^{\prime
}(y_{2}^{j};\kappa^{-1}y_{1}^{C},t_{1})\overset{y_{1}}{\circledast}%
V(\kappa^{-1}y_{1}^{B},t_{1})\nonumber\\
&  \hspace{0.05in}\qquad\qquad\quad\overset{y_{1}}{\circledast}(K_{2}^{\ast
})_{m^{-}}^{\prime}(y_{1}^{k},x^{i})+\ldots. \label{SolInt2}%
\end{align}
Let us mention that these expressions describe wave functions that emerge from
free-particle states in the remote past or future.

A short look at the identities in (\ref{DefGre1}) and (\ref{DefGre2}) should
make it obvious that the solutions in (\ref{SolInt1})-(\ref{SolInt2}%
)\ correspond to the expansions
\begin{align}
&  (G_{1})_{m^{+}}^{\prime}(z^{i},x^{j})=\,(K_{1})_{m^{+}}^{\prime}%
(z^{i};x^{A},q^{-\zeta}t_{x})\nonumber\\
&  \quad+\,\text{i}^{-1}\int_{-\infty}^{+\infty}dt_{1}\int_{-\infty}^{+\infty
}d_{1}^{n}y_{1}\,(K_{1})_{m^{+}}^{\prime}(z^{i};y_{1}^{B},q^{-\zeta}%
t_{1})\overset{y_{1}}{\circledast}V(y_{1}^{k})\nonumber\\
&  \qquad\qquad\qquad\overset{y_{1}}{\circledast}(K_{1})_{m^{+}}^{\prime
}(y_{1}^{l};x^{A},q^{-\zeta}t_{x})\nonumber\\
&  \quad+\,\text{i}^{-2}\int_{-\infty}^{+\infty}dt_{1}\int_{-\infty}^{+\infty
}d_{1}^{n}y_{1}\,\int_{-\infty}^{+\infty}dt_{2}\int_{-\infty}^{+\infty}%
d_{1}^{n}y_{2}\,(K_{1})_{m^{+}}^{\prime}(z^{i};y_{2}^{C},q^{-\zeta}%
t_{2})\nonumber\\
&  \qquad\qquad\qquad\overset{y_{2}}{\circledast}V(y_{2}^{k})\overset{y_{2}%
}{\circledast}(K_{1})_{m^{+}}^{\prime}(y_{2}^{l};y_{1}^{B},q^{-\zeta}%
t_{1})\overset{y_{1}}{\circledast}V(y_{1}^{r})\nonumber\\
&  \qquad\qquad\qquad\overset{y_{1}}{\circledast}(K_{1})_{m^{+}}^{\prime
}(y_{1}^{s};x^{A},q^{-\zeta}t_{x})+\ldots,\label{ExpGre1}\\[0.1in]
&  (G_{2})_{m^{-}}(x^{i},z^{j})=\,(K_{2})_{m^{-}}(x^{A},q^{\zeta}t_{x}%
;z^{j})\nonumber\\
&  \quad+\,\text{i}^{-1}\int_{-\infty}^{+\infty}dt_{1}\int\nolimits_{-\infty
}^{+\infty}d_{2}^{n}y_{1}\,(K_{2})_{m^{-}}(x^{A},q^{\zeta}t_{x};y_{1}%
^{k})\overset{y_{1}}{\circledast}V(y_{1}^{l})\nonumber\\
&  \qquad\qquad\qquad\overset{y_{1}}{\circledast}(K_{2})_{m^{-}}(y_{1}%
^{B},q^{\zeta}t_{1};z^{j})\nonumber\\
&  \quad+\,\text{i}^{-2}\int_{-\infty}^{+\infty}dt_{1}\int\nolimits_{-\infty
}^{+\infty}d_{2}^{n}y_{1}\,\int_{-\infty}^{+\infty}dt_{2}\int
\nolimits_{-\infty}^{+\infty}d_{2}^{n}y_{2}\,(K_{2})_{m^{-}}(x^{A},q^{\zeta
}t_{x};y_{1}^{k})\nonumber\\
&  \qquad\qquad\qquad\overset{y_{1}}{\circledast}V(y_{1}^{l})\overset{y_{1}%
}{\circledast}(K_{2})_{m^{-}}(y_{1}^{B},q^{\zeta}t_{1};y_{2}^{r}%
)\overset{y_{2}}{\circledast}V(y_{2}^{s})\nonumber\\
&  \qquad\qquad\qquad\overset{y_{2}}{\circledast}(K_{2})_{m^{-}}(y_{2}%
^{C},q^{\zeta}t_{2};z^{j})+\ldots,
\end{align}
and%
\begin{align}
&  (G_{1}^{\ast})_{m^{+}}(x^{i},z^{j})=\,(K_{1}^{\ast})_{m^{+}}(x^{i}%
,z^{j})\nonumber\\
&  \hspace{0.09in}+\,\text{i}^{-1}\int_{-\infty}^{+\infty}dt_{1}\int_{-\infty
}^{+\infty}d_{1}^{n}y_{1}\,(K_{1}^{\ast})_{m^{+}}(x^{i},y_{1}^{j}%
)\overset{y_{1}}{\circledast}V(\kappa y_{1}^{A},t_{_{1}})\nonumber\\
&  \qquad\qquad\qquad\overset{y_{1}}{\circledast}(K_{1}^{\ast})_{m^{+}}(\kappa
y_{1}^{B},t_{1};z^{j})\nonumber\\
&  \hspace{0.09in}+\,\text{i}^{-2}\int_{-\infty}^{+\infty}dt_{1}\int_{-\infty
}^{+\infty}d_{1}^{n}y_{1}\,\int_{-\infty}^{+\infty}dt_{2}\int_{-\infty
}^{+\infty}d_{1}^{n}y_{2}\,(K_{1}^{\ast})_{m^{+}}(x^{i},y_{1}^{k})\nonumber\\
&  \qquad\qquad\qquad\overset{y_{1}}{\circledast}V(\kappa y_{1}^{A},t_{_{1}%
})\overset{y_{1}}{\circledast}(K_{1}^{\ast})_{m^{+}}(\kappa y_{1}^{B}%
,t_{1};y_{2}^{l})\overset{y_{2}}{\circledast}V(\kappa y_{2}^{C},t_{2}%
)\nonumber\\
&  \qquad\qquad\qquad\overset{y_{2}}{\circledast}(K_{1}^{\ast})_{m^{+}}(\kappa
y_{2}^{D},t_{2};z^{j})+\ldots,\\[0.1in]
&  (G_{2}^{\ast})_{m^{-}}^{\prime}(z^{i},x^{j})=\,(K_{2}^{\ast})_{m^{-}%
}^{\prime}(z^{i},x^{j})\nonumber\\
&  \hspace{0.09in}+\,\text{i}^{-1}\int_{-\infty}^{+\infty}dt_{1}%
\int\nolimits_{-\infty}^{+\infty}d_{2}^{n}y_{1}\,(K_{2}^{\ast})_{m}^{\prime
}(z^{i};\kappa^{-1}y_{1}^{A},t_{1})\overset{y_{1}}{\circledast}V(\kappa
^{-1}y_{1}^{B},t_{1})\nonumber\\
&  \qquad\qquad\qquad\overset{y_{1}}{\circledast}(K_{2}^{\ast})_{m^{-}%
}^{\prime}(y_{1}^{k},x^{j})\nonumber\\
&  \hspace{0.09in}+\,\text{i}^{-2}\int_{-\infty}^{+\infty}dt_{1}%
\int\nolimits_{-\infty}^{+\infty}d_{2}^{n}y_{1}\,\int_{-\infty}^{+\infty
}dt_{2}\int\nolimits_{-\infty}^{+\infty}d_{2}^{n}y_{2}\,(K_{2}^{\ast})_{m^{-}%
}^{\prime}(z^{i};\kappa^{-1}y_{2}^{A},t_{2})\nonumber\\
&  \qquad\qquad\qquad\overset{y_{2}}{\circledast}V(\kappa^{-1}y_{2}^{B}%
,t_{2})\overset{y_{2}}{\circledast}(K_{2}^{\ast})_{m^{-}}^{\prime}(y_{2}%
^{k};\kappa^{-1}y_{1}^{C},t_{1})\overset{y_{1}}{\circledast}V(\kappa^{-1}%
y_{1}^{D},t_{1})\nonumber\\
&  \qquad\qquad\qquad\overset{y_{1}}{\circledast}(K_{2}^{\ast})_{m^{-}%
}^{\prime}(y_{1}^{l},x^{j})+\ldots\,. \label{ExpGre2}%
\end{align}

One should also notice that for each geometry we considered either retarded or
advanced Green's functions. This correspondence is a consequence of the choice
of boundary conditions in (\ref{ConBoun1}) and (\ref{ConBoun2}). If we change
these boundary conditions, the free-particle Green's functions in the above
expressions have to be substituted by their counterparts that propagate wave
functions oppositely in time.

In what follows it is necessary to know the conjugation properties of the
Green's functions introduced in (\ref{DefGre1}) and (\ref{DefGre2}) $(i=1,2)$:%
\begin{align}
\overline{(G_{i})_{m^{\pm}}(x^{k},z^{l})}  &  =(\tilde{G}_{i})_{m^{\pm}%
}^{\prime}(z^{l},x^{k}),\nonumber\\
\overline{(G_{i}^{\ast})_{m^{\pm}}(x^{k},z^{l})}  &  =(\tilde{G}_{i}^{\ast
})_{m^{\pm}}^{\prime}(z^{l},x^{k}),\label{GreCon1}\\[0.1in]
\overline{(G_{i})_{m^{\pm}}^{\prime}(z^{k},x^{l})}  &  =(\tilde{G}%
_{i})_{m^{\pm}}^{\prime}(x^{l},z^{k}),\nonumber\\
\overline{(G_{i}^{\ast})_{m^{\pm}}^{\prime}(z^{k},x^{l})}  &  =(\tilde{G}%
_{i}^{\ast})_{m^{\pm}}^{\prime}(x^{l},z^{k}). \label{GreCon2}%
\end{align}
To understand these relationship one first has to realize that the Green's
functions with a tilde obey relations in which the free-particle propagators
are substituted by those with a tilde. With this observation and the results
of (\ref{ConProp1}) and (\ref{ConProp2}) at hand the above conjugation
properties follow from conjugating the identities in (\ref{ExpGre1}%
)-(\ref{ExpGre2}).

\subsection{S-matrices and transition probabilities}

In this subsection we would like to write down probability amplitudes for
finding a certain free-particle state after the scattering process took place.
As we know, these amplitudes establish the S-matrix. In our formalism the
S-matrices for the different geometries should become%
\begin{align}
(S_{2})_{-}(\phi,\psi)  &  =\lim_{t\rightarrow-\infty}\big \langle(\phi
_{2}^{\ast})_{m}(x^{A},-t_{x}),(\psi_{2})_{m^{-}}(x^{B},t_{x}%
)\big \rangle_{2,x}\nonumber\\
&  =\lim_{t_{x}\rightarrow-\infty}\lim_{t_{y}\rightarrow\infty}\int
\nolimits_{-\infty}^{+\infty}d_{2}^{n}x\int\nolimits_{-\infty}^{+\infty}%
d_{2}^{n}y\,\overline{(\phi_{2}^{\ast})_{m}(x^{A},-t_{x})}\nonumber\\
&  \qquad\qquad\overset{x}{\circledast}(G_{2})_{m^{-}}(x^{B},t_{x};y^{C}%
,t_{y})\overset{y}{\circledast}(\phi_{2})_{m}(y^{D},q^{\zeta}t_{y}%
),\label{DefSMa0}\\[0.08in]
(S_{1}^{\ast})_{+}(\phi,\psi)  &  =\lim_{t\rightarrow+\infty}\big \langle(\phi
_{1})_{m}(x^{A},-t_{x}),(\psi_{1}^{\ast})_{m^{+}}(x^{B},t_{x}%
)\big \rangle_{1,x}\nonumber\\
&  =\lim_{t_{x}\rightarrow+\infty}\lim_{t_{y}\rightarrow-\infty}%
\int\nolimits_{-\infty}^{+\infty}d_{1}^{n}x\int\nolimits_{-\infty}^{+\infty
}d_{1}^{n}y\,\overline{(\phi_{1})_{m}(x^{A},-t_{x})}\nonumber\\
&  \qquad\qquad\overset{x}{\circledast}(G_{1}^{\ast})_{m^{+}}(x^{B}%
,t_{x};y^{C},t_{y})\overset{y}{\circledast}(\phi_{1}^{\ast})_{m}(\kappa
y^{D},t_{y}), \label{DefSMa1}%
\end{align}
and%
\begin{align}
(S_{1})_{+}^{\prime}(\phi,\psi)  &  =\lim_{t_{x}\rightarrow+\infty
}\big \langle(\psi_{1})_{m^{+}}^{\prime}(x^{A},t_{x}),(\phi_{1}^{\ast}%
)_{m}^{\prime}(x^{B},-t_{x})\big \rangle_{1,x}^{\prime}\nonumber\\
&  =\lim_{t_{x}\rightarrow+\infty}\lim_{t_{y}\rightarrow-\infty}%
\int\nolimits_{-\infty}^{+\infty}d_{1}^{n}x\int\nolimits_{-\infty}^{+\infty
}d_{1}^{n}y\,(\phi_{1})_{m}^{\prime}(y^{C},q^{-\zeta}t_{y})\nonumber\\
&  \qquad\qquad\overset{y}{\circledast}(G_{1})_{m^{+}}^{\prime}(y^{D}%
,t_{y};x^{A},t_{x})\overset{x}{\circledast}\overline{(\phi_{1}^{\ast}%
)_{m}^{\prime}(x^{B},-t_{x})},\\[0.08in]
(S_{2}^{\ast})_{-}^{\prime}(\phi,\psi)  &  =\lim_{t\rightarrow-\infty
}\big \langle(\psi_{2}^{\ast})_{m^{-}}^{\prime}(x^{A},t_{x}),(\phi_{2}%
)_{m}^{\prime}(x^{B},-t_{x})\big \rangle_{2,x}^{\prime}\nonumber\\
&  =\lim_{t_{x}\rightarrow-\infty}\lim_{t_{y}\rightarrow+\infty}%
\int\nolimits_{-\infty}^{+\infty}d_{2}^{n}x\int\nolimits_{-\infty}^{+\infty
}d_{2}^{n}y\,(\phi_{2}^{\ast})_{m}^{\prime}(\kappa^{-1}y^{C},t_{y})\nonumber\\
&  \qquad\qquad\overset{y}{\circledast}(G_{2}^{\ast})_{m^{-}}^{\prime}%
(y^{D},t_{y};x^{A},t_{x})\overset{x}{\circledast}\overline{(\phi_{2}%
)_{m}^{\prime}(x^{B},-t_{x})}. \label{DefSMa3}%
\end{align}

Notice that our S-matrix elements were formulated by means of the sesquilinear
forms introduced in part I of this article. Due to the relations in
(\ref{DefGre1}) and (\ref{DefGre2}) the S-matrix elements can be expressed by
Green's functions. In this manner, our S-matrix elements become dependent on
two free-particle wave functions, one for the incoming particle and another
one for the outcoming particle. It is important to realize that the wave
function for the incoming particle refers\ to a geometry being different from
that for the outcoming particle. However, wave functions of different
geometries can move into different directions of time. Finally, one should
notice that we have assigned a minus sign to the time argument of the wave
function of the outgoing particle. This was done to guarantee that the
free-particle wave functions involved in one and the same S-matrix element
describe particles moving into the same direction in time

Next, we would like to consider the conjugation properties of S-matrix
elements. In this respect, we have%
\begin{align}
\overline{(S_{2})_{-}(\phi,\psi)}  &  =(\tilde{S}_{2})_{-}^{\prime}(\phi
,\psi),\nonumber\\
\overline{(S_{1})_{+}^{\prime}(\phi,\psi)}  &  =(\tilde{S}_{1})_{+}^{\prime
}(\phi,\psi),\label{KonSma1}\\[0.1in]
\overline{(S_{1}^{\ast})_{+}(\phi,\psi)}  &  =(\tilde{S}_{1}^{\ast}%
)_{+}^{\prime}(\phi,\psi),\nonumber\\
\overline{(S_{2}^{\ast})_{-}^{\prime}(\phi,\psi)}  &  =(\tilde{S}_{2}^{\ast
})_{-}(\phi,\psi), \label{KonSma2}%
\end{align}
where%
\begin{align}
(\tilde{S}_{1})_{+}(\phi,\psi)  &  =\lim_{t_{x}\rightarrow+\infty}\lim
_{t_{y}\rightarrow-\infty}\int\nolimits_{-\infty}^{+\infty}d_{1}^{n}%
x\int\nolimits_{-\infty}^{+\infty}d_{1}^{n}y\,\overline{(\phi_{1}^{\ast}%
)_{m}(x^{A},-t_{x})}\nonumber\\
&  \qquad\qquad\overset{x}{\circledast}(\tilde{G}_{1})_{m^{+}}(x^{B}%
,t_{x};y^{C},t_{y})\overset{y}{\circledast}(\phi_{1})_{m}(y^{D},q^{-\zeta
}t_{y}),\\[0.08in]
(\tilde{S}_{2}^{\ast})_{-}(\phi,\psi)  &  =\lim_{t_{x}\rightarrow-\infty}%
\lim_{t_{y}\rightarrow+\infty}\int\nolimits_{-\infty}^{+\infty}d_{2}^{n}%
x\int\nolimits_{-\infty}^{+\infty}d_{2}^{n}y\,\overline{(\phi_{2})_{m}%
(x^{A},-t_{x})}\nonumber\\
&  \qquad\qquad\overset{x}{\circledast}(\tilde{G}_{2}^{\ast})_{m^{-}}%
(x^{B},t_{x};y^{C},t_{y})\overset{y}{\circledast}(\phi_{2}^{\ast})_{m}%
(\kappa^{-1}y^{D},t_{y}),
\end{align}
and%
\begin{align}
(\tilde{S}_{2})_{-}^{\prime}(\phi,\psi)  &  =\lim_{t_{x}\rightarrow-\infty
}\lim_{t_{y}\rightarrow+\infty}\int\nolimits_{-\infty}^{+\infty}d_{2}^{n}%
x\int\nolimits_{-\infty}^{+\infty}d_{2}^{n}y\,(\phi_{2})_{m}^{\prime}%
(y^{A},q^{\zeta}t_{y})\nonumber\\
&  \qquad\qquad\overset{y}{\circledast}(\tilde{G}_{2})_{m^{-}}^{\prime}%
(y^{B},t_{y};x^{C},t_{x})\overset{x}{\circledast}\overline{(\phi_{2}^{\ast
})_{m}^{\prime}(x^{D},-t_{x})},\\[0.08in]
(\tilde{S}_{1}^{\ast})_{+}^{\prime}(\phi,\psi)  &  =\lim_{t_{x}\rightarrow
+\infty}\lim_{t_{y}\rightarrow-\infty}\int\nolimits_{-\infty}^{+\infty}%
d_{1}^{n}x\int\nolimits_{-\infty}^{+\infty}d_{1}^{n}y\,(\phi_{1}^{\ast}%
)_{m}^{\prime}(\kappa y^{A},t_{y})\nonumber\\
&  \qquad\qquad\overset{y}{\circledast}(\tilde{G}_{1}^{\ast})_{m^{+}}^{\prime
}(y^{B},t_{y};x^{C},t_{x})\overset{x}{\circledast}\overline{(\phi_{1}%
)_{m}^{\prime}(x^{D},-t_{x})}.
\end{align}

The relations in (\ref{KonSma1}) and (\ref{KonSma2}) can be proved in the
following manner:%
\begin{align}
\overline{(S_{2})_{-}(\phi,\psi)}  &  =\lim_{t_{x}\rightarrow-\infty}%
\overline{\big \langle(\phi_{2}^{\ast})_{m}(x^{A},-t_{x}),(\psi_{2})_{m}%
(x^{B},t_{x})\big \rangle_{i,x}}\nonumber\\
&  =\lim_{t_{x}\rightarrow-\infty}\lim_{t_{y}\rightarrow+\infty}%
\int\nolimits_{-\infty}^{+\infty}d_{2}^{n}x\int\nolimits_{-\infty}^{+\infty
}d_{2}^{n}y\,\overline{(\phi_{i})_{m}(y^{D},q^{\zeta}t_{y})}\nonumber\\
&  \qquad\qquad\overset{y}{\circledast}\overline{(G_{2})_{m^{-}}(x^{B}%
,t_{x};y^{C},t_{y})}\overset{x}{\circledast}(\phi_{2}^{\ast})_{m}(x^{A}%
,-t_{x})\nonumber\\
&  =\lim_{t_{x}\rightarrow-\infty}\lim_{t_{y}\rightarrow+\infty}%
\int\nolimits_{-\infty}^{+\infty}d_{2}^{n}x\int\nolimits_{-\infty}^{+\infty
}d_{2}^{n}y\,(\phi_{2})_{m}(y^{D},q^{\zeta}t_{y})\nonumber\\
&  \qquad\qquad\overset{x}{\circledast}(\tilde{G}_{2})_{m^{-}}^{\prime}%
(y^{C},t_{y};x^{B},t_{x})\overset{x}{\circledast}\overline{(\phi_{2}^{\ast
})_{m}^{\prime}(x^{A},-t_{x})}\nonumber\\
&  =(\tilde{S}_{2})_{-}^{\prime}(\phi,\psi).
\end{align}
The first and the second step in the above calculation make use of the
identities in (\ref{DefSMa0}). The third step applies the results of
(\ref{GreCon1}) together with the identifications (see part II of the article)%
\begin{equation}
\overline{(\phi_{i})_{m}(x^{k})}=(\phi_{i})_{m}^{\prime}(x^{k}),\qquad
\overline{(\phi_{i}^{\ast})_{m}(x^{k})}=(\phi_{i}^{\ast})_{m}^{\prime}(x^{k}).
\end{equation}

Now, we are in a position to introduce transition probabilities. It seems to
be reasonable to define these quantities by%
\begin{align}
(\omega_{2})_{-}(\phi,\psi)  &  \equiv\overline{(S_{2})_{-}(\phi,\psi)}%
\cdot(S_{2})_{-}(\phi,\psi),\nonumber\\
(\omega_{1}^{\ast})_{+}(\phi,\psi)  &  \equiv\overline{(S_{1}^{\ast})_{+}%
(\phi,\psi)}\cdot(S_{1}^{\ast})_{+}(\phi,\psi),\label{TraPro1}\\[0.1in]
(\omega_{1})_{+}^{\prime}(\phi,\psi)  &  \equiv(S_{1})_{+}^{\prime}(\phi
),\psi)\cdot\overline{(S_{1})_{+}^{\prime}(\phi,\psi)},\nonumber\\
(\omega_{2}^{\ast})_{-}^{\prime}(\phi,\psi)  &  \equiv(S_{2}^{\ast}%
)_{-}^{\prime}(\phi,\psi)\cdot\overline{(S_{2}^{\ast})_{-}^{\prime}(\phi
,\psi)}. \label{TraPro2}%
\end{align}
From their very definition it follows that the transition probabilities are real.

Alternatively, we can introduce transition probabilities by%
\begin{align}
(\tilde{\omega}_{1})_{-}(\phi,\psi)  &  \equiv\overline{(\tilde{S}_{1}%
)_{+}(\phi,\psi)}\cdot(\tilde{S}_{1})_{+}(\phi,\psi),\nonumber\\
(\tilde{\omega}_{2}^{\ast})_{-}(\phi,\psi)  &  \equiv\overline{(\tilde{S}%
_{2}^{\ast})_{-}(\phi,\psi)}\cdot(\tilde{S}_{2}^{\ast})_{-}(\phi
,\psi),\\[0.1in]
(\tilde{\omega}_{2})_{-}^{\prime}(\phi,\psi)  &  \equiv(\tilde{S}_{2}%
)_{-}^{\prime}(\phi,\psi)\cdot\overline{(\tilde{S}_{2})_{-}^{\prime}(\phi
,\psi)},\nonumber\\
(\tilde{\omega}_{1}^{\ast})_{+}^{\prime}(\phi,\psi)  &  \equiv(\tilde{S}%
_{1}^{\ast})_{+}^{\prime}(\phi,\psi)\cdot\overline{(\tilde{S}_{1}^{\ast}%
)_{+}^{\prime}(\phi,\psi)}.
\end{align}
They are linked to those in (\ref{TraPro1}) and (\ref{TraPro2}) via
\begin{align}
(\omega_{2})_{-}(\phi,\psi)  &  =(\tilde{\omega}_{2})_{-}^{\prime}(\phi
,\psi),\nonumber\\
(\omega_{1}^{\ast})_{+}(\phi,\psi)  &  =(\tilde{\omega}_{1}^{\ast}%
)_{+}^{\prime}(\phi,\psi),\label{ConW1}\\[0.1in]
(\omega_{1})_{+}^{\prime}(\phi,\psi)  &  =(\tilde{\omega}_{1})_{+}(\phi
,\psi),\nonumber\\
(\omega_{2}^{\ast})_{-}^{\prime}(\phi,\psi)  &  =(\tilde{\omega}_{2}^{\ast
})_{-}(\phi,\psi). \label{ConW2}%
\end{align}
These identifications follow from\ the relations in (\ref{KonSma1}) and
(\ref{KonSma2}).

Now, we come to the question what becomes of unitarity of the S-matrix in a
q-deformed setting. To answer this question it is convenient to determine the
S-matrix in an orthonormal basis of q-deformed momentum eigenfunctions.
Towards this end we take the expansions of free-particle wave functions in
terms of plane waves and insert them into the defining expressions of the
S-matrices [cf. (\ref{DefSMa0})-(\ref{DefSMa3})]. This way, we obtain, for
example,%
\begin{align}
&  (S_{2})_{-}(\phi,\psi)=\nonumber\\
&  \qquad=\lim_{t_{x}\rightarrow-\infty}\lim_{t_{y}\rightarrow+\infty}%
\int\nolimits_{-\infty}^{+\infty}d_{2}^{n}x\int\nolimits_{-\infty}^{+\infty
}d_{2}^{n}y\,\overline{(\phi_{2}^{\ast})_{m}(x^{A},-t_{x})}\nonumber\\
&  \qquad\qquad\qquad\overset{x}{\circledast}(G_{2})_{m^{-}}(x^{B},t_{x}%
;y^{C},t_{y})\overset{y}{\circledast}(\phi_{2})_{m}(y^{D},q^{\zeta}%
t_{y})\nonumber\\
&  \qquad=\lim_{t_{x}\rightarrow-\infty}\lim_{t_{y}\rightarrow+\infty}%
\int\nolimits_{-\infty}^{+\infty}d_{2}^{n}x\int\nolimits_{-\infty}^{+\infty
}d_{2}^{n}y\,(\text{vol}_{2})^{1/2}\Big (\int\nolimits_{-\infty}^{+\infty
}d_{2}^{n}p\,(c_{2}^{\ast})_{\kappa p}^{\prime}\nonumber\\
&  \qquad\qquad\qquad\overset{p}{\circledast}(\bar{u}_{R,\bar{L}})_{p,m}%
(x^{A},-t_{x})\Big )\overset{x}{\circledast}(G_{2})_{m^{-}}(x^{B},t_{x}%
;y^{C},t_{y})\nonumber\\
&  \qquad\qquad\qquad\overset{y}{\circledast}\frac{\kappa^{-n}}{(\text{vol}%
_{2})^{1/2}}\int_{-\infty}^{+\infty}d_{2}^{n}p^{\prime}\,(\bar{u}_{R,\bar{L}%
})_{\ominus_{R}p^{\prime},m}(y^{D},q^{\zeta}t_{y})\overset{y|p^{\prime}}%
{\odot}_{\hspace{-0.02in}L}(c_{2})_{\kappa^{-1}p^{\prime}}\nonumber\\
&  \qquad=\int\nolimits_{-\infty}^{+\infty}d_{2}^{n}p\,\int_{-\infty}%
^{+\infty}d_{2}^{n}p^{\prime}\,(c_{2}^{\ast})_{\kappa p}^{\prime}\overset
{p}{\circledast}[(S_{2})_{-}]_{p(\kappa p^{\prime})}\overset{p^{\prime}%
}{\circledast}(c_{2})_{p^{\prime}},
\end{align}
where we introduced as elements of the S-matrix in a momentum basis%
\begin{align}
\lbrack(S_{2})_{-}]_{pp^{\prime}}  &  =\lim_{t_{x}\rightarrow-\infty}%
\lim_{t_{y}\rightarrow+\infty}\int\nolimits_{-\infty}^{+\infty}d_{2}^{n}%
x\int\nolimits_{-\infty}^{+\infty}d_{2}^{n}y\,(\bar{u}_{R,\bar{L}}%
)_{p,m}(x^{A},-t_{x})\nonumber\\
&  \qquad\qquad\qquad\qquad\overset{x}{\circledast}(G_{2})_{m^{-}}(x^{B}%
,t_{x};y^{C},t_{y})\nonumber\\
&  \qquad\qquad\qquad\qquad\overset{y|p^{\prime}}{\odot}_{\hspace{-0.02in}%
L}(\bar{u}_{R,\bar{L}})_{\ominus_{R}p^{\prime},m}(y^{D},\kappa^{-2}q^{-\zeta
}t_{y}).
\end{align}
With the same reasonings we find for the other geometries that%
\begin{align}
&  (S_{1}^{\ast})_{+}(\phi,\psi)=\nonumber\\
&  \qquad=\int\nolimits_{-\infty}^{+\infty}d_{1}^{n}p\int\nolimits_{-\infty
}^{+\infty}d_{1}^{n}p^{\prime}\,(c_{1})_{p}^{\prime}\overset{p}{\circledast
}[(S_{1}^{\ast})_{+}]_{(\kappa^{-1}p)p^{\prime}}\overset{p^{\prime}%
}{\circledast}(c_{1}^{\ast})_{\kappa^{-1}p^{\prime}},\\[0.08in]
&  (S_{1})_{+}^{\prime}(\phi,\psi)=\nonumber\\
&  \qquad=\int\nolimits_{-\infty}^{+\infty}d_{1}^{n}p\,\int_{-\infty}%
^{+\infty}d_{1}^{n}p^{\prime}\,(c_{1})_{p}^{\prime}\overset{p}{\circledast
}[(S_{1})_{+}^{\prime}]_{(\kappa^{-1}p)p^{\prime}}\overset{p^{\prime}%
}{\circledast}(c_{1}^{\ast})_{\kappa^{-1}p^{\prime}},\\[0.08in]
&  (S_{2}^{\ast})_{-}^{\prime}(\phi,\psi)=\nonumber\\
&  \qquad=\int\nolimits_{-\infty}^{+\infty}d_{2}^{n}p\int\nolimits_{-\infty
}^{+\infty}d_{2}^{n}p^{\prime}\,(c_{2}^{\ast})_{\kappa p}^{\prime}\overset
{p}{\circledast}[(S_{2}^{\ast})_{-}^{\prime}]_{p(\kappa p^{\prime})}%
\overset{p^{\prime}}{\circledast}(c_{2})_{p^{\prime}},
\end{align}
where
\begin{align}
\lbrack(S_{1}^{\ast})_{+}]_{pp^{\prime}}  &  =\lim_{t_{x}\rightarrow+\infty
}\lim_{t_{y}\rightarrow-\infty}\int\nolimits_{-\infty}^{+\infty}d_{1}^{n}%
x\int\nolimits_{-\infty}^{+\infty}d_{1}^{n}y\,(u_{\bar{R},L})_{\ominus_{L}%
p,m}(x^{A},-\kappa^{2}t_{x})\nonumber\\
&  \qquad\qquad\qquad\qquad\overset{p|x}{\odot}_{\hspace{-0.01in}\bar{L}%
}(G_{1}^{\ast})_{m^{\pm}}(x^{B},t_{x};y^{C},t_{y})\nonumber\\
&  \qquad\qquad\qquad\qquad\overset{y}{\circledast}(u_{\bar{R},L})_{p^{\prime
},m}(y^{D},t_{y}),\\[0.08in]
\lbrack(S_{1})_{+}^{\prime}]_{pp^{\prime}}  &  =\lim_{t_{x}\rightarrow+\infty
}\lim_{t_{y}\rightarrow-\infty}\int\nolimits_{-\infty}^{+\infty}d_{1}^{n}%
x\int\nolimits_{-\infty}^{+\infty}d_{1}^{n}y\,(u_{\bar{R},L})_{\ominus_{L}%
p,m}(y^{A},\kappa^{2}q^{-\zeta}t_{y})\nonumber\\
&  \qquad\qquad\qquad\qquad\overset{p|y}{\odot}_{\hspace{-0.01in}R}%
(G_{1})_{m^{+}}^{\prime}(y^{B},t_{y};x^{C},t_{x})\nonumber\\
&  \qquad\qquad\qquad\qquad\overset{x}{\circledast}(u_{\bar{R},L})_{p^{\prime
},m}(x^{D},-t_{x}),\\[0.08in]
\lbrack(S_{2}^{\ast})_{-}^{\prime}]_{pp^{\prime}}  &  =\lim_{t_{x}%
\rightarrow-\infty}\lim_{t_{y}\rightarrow+\infty}\int\nolimits_{-\infty
}^{+\infty}d_{2}^{n}x\int\nolimits_{-\infty}^{+\infty}d_{2}^{n}y\,(\bar
{u}_{R,\bar{L}})_{p,m}(y^{A},t_{y})\nonumber\\
&  \qquad\qquad\qquad\qquad\overset{y}{\circledast}(G_{2}^{\ast})_{m^{-}%
}^{\prime}(y^{B},t_{y};x^{C},t_{x})\nonumber\\
&  \qquad\qquad\qquad\qquad\overset{x|p}{\odot}_{\hspace{-0.01in}\bar{R}}%
(\bar{u}_{R,\bar{L}})_{\ominus_{R}p^{\prime},m}(y^{B},-\kappa^{-2}t_{y}).
\end{align}

\begin{figure}
[tbh]
\begin{center}
\includegraphics[
height=3.8488in,
width=4.0045in
]%
{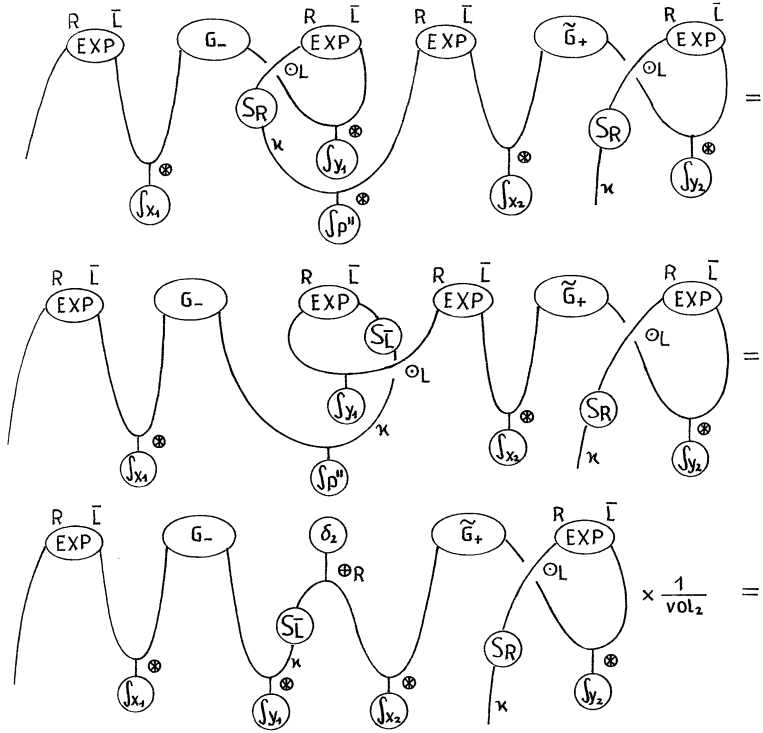}%
\caption{First part of a diagrammatic proof corresponding to (\ref{ImDaGrInv1}%
).}%
\label{Fig1}%
\end{center}
\end{figure}

\begin{figure}
[tbhtbh]
\begin{center}
\includegraphics[
height=3.4139in,
width=3.8422in
]%
{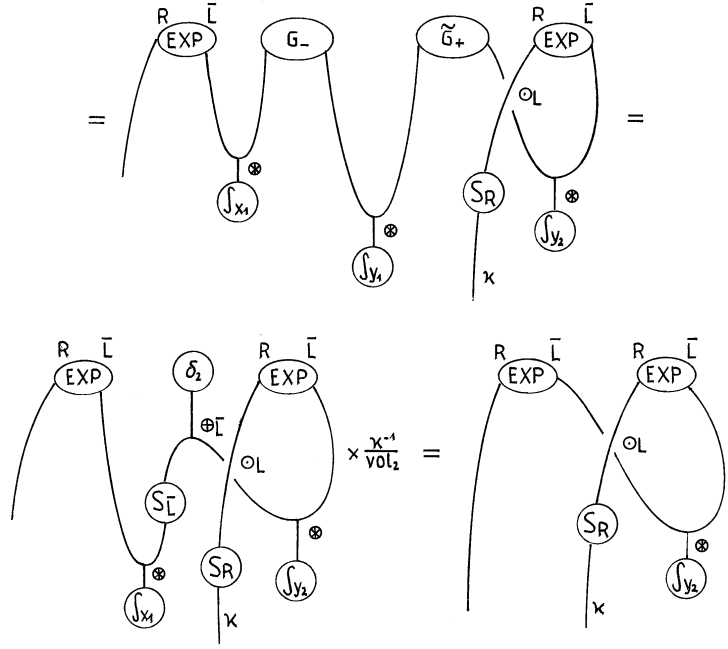}%
\caption{Second part of a diagrammatic proof corresponding to
(\ref{ImDaGrInv1}).}%
\label{Fig2}%
\end{center}
\end{figure}

As next step we calculate products of S-matrices in a momentum basis.
Concretely, we are interested in products like the following one:%
\begin{align}
&  \int_{-\infty}^{+\infty}d_{2}^{n}p^{\prime\prime}\,[(S_{2})_{-}]_{p(\kappa
p^{\prime\prime})}\overset{p^{\prime\prime}}{\circledast}[(\tilde{S}_{2}%
)_{+}]_{p^{\prime\prime}(\kappa p^{\prime})}=\nonumber\\
&  =\lim_{t\rightarrow-\infty}\lim_{t^{\prime}\rightarrow+\infty}\int
_{-\infty}^{+\infty}d_{2}^{n}p^{\prime\prime}\int\nolimits_{-\infty}^{+\infty
}d_{2}^{n}x_{1}\int\nolimits_{-\infty}^{+\infty}d_{2}^{n}y_{1}\,(\bar
{u}_{R,\bar{L}})_{p,m}(x_{1}^{A},-t)\nonumber\\
&  \qquad\qquad\overset{x_{1}}{\circledast}(G_{2})_{m^{-}}(x_{1}^{B}%
,t;y_{1}^{C},t^{\prime})\overset{y_{1}|p^{\prime\prime}}{\odot}_{\hspace
{-0.05in}L}(\bar{u}_{R,\bar{L}})_{\ominus_{R}(\kappa p^{\prime\prime}%
),m}(y_{1}^{D},\kappa^{-2}q^{\zeta}t^{\prime})\nonumber\\
&  \qquad\qquad\overset{p^{\prime\prime}}{\circledast}\int\nolimits_{-\infty
}^{+\infty}d_{2}^{n}x_{2}\int\nolimits_{-\infty}^{+\infty}d_{2}^{n}%
y_{2}\,(\bar{u}_{R,\bar{L}})_{p^{\prime\prime},m}(x_{2}^{E},-t^{\prime
})\nonumber\\
&  \qquad\qquad\overset{x_{2}}{\circledast}(\tilde{G}_{2})_{m^{+}}(x_{2}%
^{F},t^{\prime};y_{2}^{G},t)\overset{y_{2}|p^{\prime}}{\odot}_{\hspace
{-0.04in}L}(\bar{u}_{R,\bar{L}})_{\ominus_{R}(\kappa p^{\prime}),m}(y_{2}%
^{H},\kappa^{-2}q^{\zeta}t)\nonumber\\
&  =\lim_{t\rightarrow-\infty}\lim_{t^{\prime}\rightarrow+\infty}%
\int\nolimits_{-\infty}^{+\infty}d_{2}^{n}x_{1}\int\nolimits_{-\infty
}^{+\infty}d_{2}^{n}y_{1}\,\frac{1}{\text{vol}_{2}}(\bar{u}_{R,\bar{L}}%
)_{p,m}(x_{1}^{A},-t)\nonumber\\
&  \qquad\quad\overset{x_{1}}{\circledast}(G_{2})_{m^{-}}(x_{1}^{B}%
,t;y_{1}^{C},t^{\prime})\overset{y_{1}}{\circledast}\int\nolimits_{-\infty
}^{+\infty}d_{2}^{n}x_{2}\,\delta_{2}^{n}((\ominus_{\bar{L}}(\kappa y_{1}%
^{D})\oplus_{\bar{L}}x_{2}^{E})\nonumber\\
&  \qquad\quad\overset{x_{2}}{\circledast}\int\nolimits_{-\infty}^{+\infty
}d_{2}^{n}y_{2}\,(\tilde{G}_{2})_{m^{+}}(x_{2}^{F},t^{\prime};y_{2}%
^{G},t)\overset{y_{2}|p^{\prime}}{\odot}_{\hspace{-0.04in}L}(\bar{u}%
_{R,\bar{L}})_{\ominus_{R}(\kappa p^{\prime}),m}(y_{2}^{H},\kappa^{-2}%
q^{\zeta}t)\nonumber\\
&  =\lim_{t\rightarrow-\infty}\lim_{t^{\prime}\rightarrow+\infty}%
\int\nolimits_{-\infty}^{+\infty}d_{2}^{n}x_{1}\int\nolimits_{-\infty
}^{+\infty}d_{2}^{n}y_{1}\int\nolimits_{-\infty}^{+\infty}d_{2}^{n}%
y_{2}\,(\bar{u}_{R,\bar{L}})_{p,m}(x_{1}^{A},-t)\nonumber\\
&  \qquad\quad\overset{x_{1}}{\circledast}(G_{2})_{m^{-}}(x_{1}^{B}%
,t;y_{1}^{C},t^{\prime})\overset{y_{1}}{\circledast}(\tilde{G}_{2})_{m^{+}%
}(x_{2}^{D},t^{\prime};y_{2}^{E},t)\nonumber\\
&  \qquad\quad\quad\overset{y_{2}|p^{\prime}}{\odot}_{\hspace{-0.04in}L}%
(\bar{u}_{R,\bar{L}})_{\ominus_{R}(\kappa p^{\prime}),m}(y_{2}^{F},\kappa
^{-2}q^{\zeta}t)\nonumber\\
&  =\frac{\kappa^{-n}}{\text{vol}_{2}}\int\nolimits_{-\infty}^{+\infty}%
d_{2}^{n}x_{1}\int\nolimits_{-\infty}^{+\infty}d_{2}^{n}y_{2}\,(\bar
{u}_{R,\bar{L}})_{p,m}(x_{1}^{A},-t)\nonumber\\
&  \qquad\quad\overset{x_{1}}{\circledast}\delta_{2}^{n}((\ominus_{\bar{L}%
}\,x_{1}^{B})\oplus_{\bar{L}}(\kappa^{-1}y_{2}^{C}))\overset{y_{2}|p^{\prime}%
}{\odot}_{\hspace{-0.04in}L}(\bar{u}_{R,\bar{L}})_{\ominus_{R}(\kappa
p^{\prime}),m}(y_{2}^{D},\kappa^{-2}q^{\zeta}t)\nonumber\\
&  =\int\nolimits_{-\infty}^{+\infty}d_{2}^{n}x_{1}\,(\bar{u}_{R,\bar{L}%
})_{p,m}(x_{1}^{A},-t)\overset{x_{1}|p^{\prime}}{\odot}_{\hspace{-0.04in}%
L}(\bar{u}_{R,\bar{L}})_{\ominus_{R}(\kappa p^{\prime}),m}(x_{1}^{B}%
,\kappa^{-2}q^{\zeta}t)\nonumber\\
&  =(\text{vol}_{2})^{-1}\delta_{2}^{n}(p^{A}\oplus_{R}(\ominus_{R}\,\kappa
p^{\prime B})).\label{ImDaGrInv1}%
\end{align}
For the first step we plug in the expressions of the S-matrices in a momentum
basis. Then we identify the completeness relation of plane waves, i.e.%
\begin{align}
&  \int_{-\infty}^{+\infty}d_{2}^{n}p\text{\thinspace}(\bar{u}_{R,\bar{L}%
})_{\ominus_{R}p,m}(y^{B},q^{\zeta}t)\overset{y|p}{\odot}_{\hspace
{-0.01in}\bar{R}}(\bar{u}_{R,\bar{L}})_{p,m}(x^{A},-t)=\nonumber\\
&  \hspace{0.4in}\hspace{0.4in}=\,(\text{vol}_{2})^{-1}\delta_{2}^{n}%
((\ominus_{\bar{L}}\,y^{B})\oplus_{\bar{L}}x^{A}).
\end{align}
In this manner, we get a q-deformed delta function. Exploiting its
characteristic property
\begin{equation}
\int\nolimits_{-\infty}^{+\infty}d_{2}^{n}x\,\delta_{2}^{n}((\ominus_{\bar{L}%
}\,y^{A})\oplus_{\bar{L}}x^{B})\overset{x}{\circledast}f(x^{C})=\text{vol}%
_{2}\,f(\kappa^{-1}y^{A}),\label{DeltProAlg}%
\end{equation}
one integration vanishes. Then the fourth step uses the identities in
(\ref{GreeInv1}) and leads to another q-deformed delta function. For this
reason a further integral disappears. The last step is nothing other than
orthonormality of plane waves. For these reasonings to become more clear we
additionally added their diagrammatic version in Figs. \ref{Fig1} and
\ref{Fig2} (for an introduction into this subject we refer the reader to Ref.
\cite{Maj93-Int}).%

After repeating these steps for the other geometries we should end up with the
following list of relations:%
\begin{align}
&  \int_{-\infty}^{+\infty}d_{2}^{n}p^{\prime\prime}\,[(S_{2})_{-}]_{p(\kappa
p^{\prime\prime})}\overset{p^{\prime\prime}}{\circledast}[(\tilde{S}_{2}%
)_{+}]_{p^{\prime\prime}(\kappa p^{\prime})}=\nonumber\\
&  \qquad=\int_{-\infty}^{+\infty}d_{2}^{n}p^{\prime\prime}\,[(\tilde{S}%
_{2})_{+}]_{p(\kappa p^{\prime\prime})}\overset{p^{\prime\prime}}{\circledast
}[(S_{2})_{-}]_{p^{\prime\prime}(\kappa p^{\prime})}\nonumber\\
&  \qquad=(\text{vol}_{2})^{-1}\delta_{2}^{n}(p^{A}\oplus_{R}(\ominus
_{R}\,\kappa p^{\prime B})),\label{ImDaGrInv1N}\\[0.1in]
&  \int_{-\infty}^{+\infty}d_{1}^{n}p^{\prime\prime}\,[(S_{1}^{\ast}%
)_{+}]_{(\kappa^{-1}p)p^{\prime\prime}}\overset{p^{\prime\prime}}{\circledast
}[(\tilde{S}_{1}^{\ast})_{-}]_{(\kappa^{-1}p^{\prime\prime})p^{\prime}%
}=\nonumber\\
&  \qquad=\int_{-\infty}^{+\infty}d_{1}^{n}p^{\prime\prime}\,[(\tilde{S}%
_{1}^{\ast})_{-}]_{(\kappa^{-1}p)p^{\prime\prime}}\overset{p^{\prime\prime}%
}{\circledast}[(S_{1}^{\ast})_{+}]_{(\kappa^{-1}p^{\prime\prime})p^{\prime}%
}\nonumber\\
&  \qquad=(\text{vol}_{1})^{-1}\delta_{1}^{n}((\ominus_{L}\,\kappa^{-1}%
p^{A})\oplus_{L}p^{\prime B}),
\end{align}
and%
\begin{align}
&  \int_{-\infty}^{+\infty}d_{1}^{n}p^{\prime\prime}\,[(S_{1})_{+}^{\prime
}]_{(\kappa^{-1}p)p^{\prime\prime}}\overset{p^{\prime\prime}}{\circledast
}[(\tilde{S}_{1})_{-}^{\prime}]_{(\kappa^{-1}p^{\prime\prime})p^{\prime}%
}=\nonumber\\
&  \qquad=\int_{-\infty}^{+\infty}d_{1}^{n}p^{\prime\prime}\,[(\tilde{S}%
_{1})_{-}^{\prime}]_{(\kappa^{-1}p)p^{\prime\prime}}\overset{p^{\prime\prime}%
}{\circledast}[(S_{1})_{+}^{\prime}]_{(\kappa^{-1}p^{\prime\prime})p^{\prime}%
}\nonumber\\
&  \qquad=(\text{vol}_{1})^{-1}\delta_{1}^{n}((\ominus_{L}\,\kappa^{-1}%
p^{A})\oplus_{L}p^{\prime B}),\\[0.1in]
&  \int_{-\infty}^{+\infty}d_{2}^{n}p^{\prime\prime}\,[(S_{2}^{\ast}%
)_{-}^{\prime}]_{p(\kappa p^{\prime\prime})}\overset{p^{\prime\prime}%
}{\circledast}[(\tilde{S}_{2}^{\ast})_{+}^{\prime}]_{p^{\prime\prime}(\kappa
p^{\prime})}=\nonumber\\
&  \qquad=\int_{-\infty}^{+\infty}d_{2}^{n}p^{\prime\prime}\,[(\tilde{S}%
_{2}^{\ast})_{+}^{\prime}]_{p(\kappa p^{\prime\prime})}\overset{p^{\prime
\prime}}{\circledast}[(S_{2}^{\ast})_{-}^{\prime}]_{p^{\prime\prime}(\kappa
p^{\prime})}\nonumber\\
&  \qquad=(\text{vol}_{2})^{-1}\delta_{2}^{n}(p^{A}\oplus_{R}(\ominus
_{R}\,\kappa p^{\prime B})). \label{ImDaGrInv2N}%
\end{align}
The above relations tell us that in a momentum basis the S-matrices are
invertible. For each S-matrix the corresponding inverse can be read off from
the results in (\ref{ImDaGrInv1N})-(\ref{ImDaGrInv2N}).

Interestingly, the inverse of a given S-matrix can be identified with the
Hermitian conjugate of another S-matrix. In this sense, we have%
\begin{align}
\overline{\lbrack(S_{2})_{-}]_{pp^{\prime}}}  &  =[(\tilde{S}_{2})_{-}%
^{\prime}]_{p^{\prime}p},\qquad\overline{[(S_{1}^{\ast})_{+}]_{pp^{\prime}}%
}=[(\tilde{S}_{1}^{\ast})_{+}^{\prime}]_{p^{\prime}p},\nonumber\\
\overline{\lbrack(S_{1})_{+}^{\prime}]_{pp^{\prime}}}  &  =[(\tilde{S}%
_{1})_{+}]_{p^{\prime}p},\qquad\overline{[(S_{2}^{\ast})_{-}^{\prime
}]_{pp^{\prime}}}=[(\tilde{S}_{2}^{\ast})_{-}]_{p^{\prime}p}. \label{ConSMa}%
\end{align}
These equalities can be readily checked if we insert the expressions for the
S-matrix elements and take into account the conjugation properties of Green's
functions and plane waves.

Using the conjugation properties in (\ref{ConSMa}) the relations in
(\ref{ImDaGrInv1N})-(\ref{ImDaGrInv2N}) can be rewritten as
\begin{align}
&  \int_{-\infty}^{+\infty}d_{2}^{n}p^{\prime\prime}\,[(S_{2})_{-}]_{p(\kappa
p^{\prime\prime})}\overset{p^{\prime\prime}}{\circledast}\overline
{[(S_{2})_{+}^{\prime}]_{p^{\prime}p^{\prime\prime}}}=\nonumber\\
&  \qquad=\int_{-\infty}^{+\infty}d_{2}^{n}p^{\prime\prime}\,\overline
{[(S_{2})_{+}^{\prime}]_{(\kappa p^{\prime\prime})p}}\overset{p^{\prime\prime
}}{\circledast}[(S_{2})_{-}]_{p^{\prime\prime}p^{\prime}}\nonumber\\
&  \qquad=(\text{vol}_{2})^{-1}\delta_{2}^{n}(p^{A}\oplus_{R}(\ominus
_{R}\,p^{\prime B})),\\[0.1in]
&  \int_{-\infty}^{+\infty}d_{1}^{n}p^{\prime\prime}\,[(S_{1}^{\ast}%
)_{+}]_{pp^{\prime\prime}}\overset{p^{\prime\prime}}{\circledast}%
\overline{[(S_{1}^{\ast})_{-}^{\prime}]_{p^{\prime}(\kappa^{-1}p^{\prime
\prime})}}=\nonumber\\
&  \qquad=\int_{-\infty}^{+\infty}d_{1}^{n}p^{\prime\prime}\,\overline
{[(S_{1}^{\ast})_{-}^{\prime}]_{p^{\prime\prime}p}}\overset{p^{\prime\prime}%
}{\circledast}[(S_{1}^{\ast})_{+}]_{(\kappa^{-1}p^{\prime\prime})p^{\prime}%
}\nonumber\\
&  \qquad=(\text{vol}_{1})^{-1}\delta_{1}^{n}((\ominus_{L}\,p^{A})\oplus
_{L}p^{\prime B}),
\end{align}
and%
\begin{align}
&  \int_{-\infty}^{+\infty}d_{1}^{n}p^{\prime\prime}\,[(S_{1})_{+}^{\prime
}]_{pp^{\prime\prime}}\overset{p^{\prime\prime}}{\circledast}\overline
{[(S_{1})_{-}]_{p^{\prime}(\kappa^{-1}p^{\prime\prime})}}=\nonumber\\
&  \qquad=\int_{-\infty}^{+\infty}d_{1}^{n}p^{\prime\prime}\,\overline
{[(S_{1})_{-}]_{p^{\prime\prime}p}}\overset{p^{\prime\prime}}{\circledast
}[(S_{1})_{+}^{\prime}]_{(\kappa^{-1}p^{\prime\prime})p^{\prime}}\nonumber\\
&  \qquad=(\text{vol}_{1})^{-1}\delta_{1}^{n}((\ominus_{L}\,p^{A})\oplus
_{L}p^{\prime B}),\\[0.1in]
&  \int_{-\infty}^{+\infty}d_{2}^{n}p^{\prime\prime}\,[(S_{2}^{\ast}%
)_{-}^{\prime}]_{p(\kappa p^{\prime\prime})}\overset{p^{\prime\prime}%
}{\circledast}\overline{[(S_{2}^{\ast})_{+}]_{p^{\prime}p^{\prime\prime}}%
}=\nonumber\\
&  \qquad=\int_{-\infty}^{+\infty}d_{2}^{n}p^{\prime\prime}\,\overline
{[(S_{2}^{\ast})_{+}]_{(\kappa p^{\prime\prime})p}}\overset{p^{\prime\prime}%
}{\circledast}[(S_{2}^{\ast})_{-}^{\prime}]_{p^{\prime\prime}p^{\prime}%
}\nonumber\\
&  \qquad=(\text{vol}_{2})^{-1}\delta_{2}^{n}(p^{A}\oplus_{R}(\ominus
_{R}\,p^{\prime B})).
\end{align}
In this manner, we arrived at identities that express unitarity of S-matrices
in the setting of q-deformation.

Finally, we would like to remind the reader of the fact that in this section
we dealt with certain geometries, only. The expressions corresponding to the
other geometries are again obtained by applying the substitutions in
(\ref{SubGeo}) to the results of this section.

\section{The interaction picture\label{IntPic}}

In this section we would like to show that our formalism allows to consider
the process of scattering from a point of view provided by the interaction
picture. Let us recall that the interaction picture is nothing other than a
reformulation of the Schr\"{o}dinger picture. Particularly, it becomes very
useful\ in describing scattering processes.

In a scattering process we usually start from free-particle states, i.e. their
time-evolution is determined by a free-particle Hamiltonian $H_{0}.$ Such a
free-particle state can be expanded in terms of momentum eigenstates and the
corresponding expansion coefficients describe to which extent the different
momentum eigenstates are populated. Notice that in our formalism momentum
eigenstates are represented by q-deformed plane waves.

If the particle interacts with a potential $V$ its wave function can again be
expanded in terms of momentum eigenfunctions, since q-deformed plane waves
establish a complete and orthonormal set of functions. Under the influence of
the potential $V$ this expansion can change in time. For this reason the
expansion coefficients now show an additional time-dependence which contains
information about the probability for a particle to be forced into a certain
momentum state.

The time dependence of the plane waves, however, can be neglected, since it
does not carry any information about the interaction. In this manner, the wave
functions describing the scattering process in the interaction picture should
become%
\begin{align}
(\Psi_{1})_{m}^{\prime}(x^{i})  &  =\frac{\kappa^{n}}{(\text{vol}_{1})^{1/2}%
}\int_{-\infty}^{+\infty}d_{1}^{n}p\,(C_{1})_{\kappa p}^{\prime}%
(t)\overset{p|x}{\odot}_{\hspace{-0.01in}R}(u_{\bar{R},L})_{\ominus_{L}%
p,m}(x^{A},t=0)\nonumber\\
&  =\frac{\kappa^{n}}{(\text{vol}_{1})^{1/2}}\big \langle(C_{1})_{\kappa
p}^{\prime}(t),(\bar{u}_{\bar{R},L})_{\ominus_{\bar{R}}p,m}(x^{A}%
,t=0)\big \rangle_{1,p}^{\prime},\label{EntInt1}\\[0.1in]
(\Psi_{2})_{m}(x^{i})  &  =\frac{\kappa^{-n}}{(\text{vol}_{2})^{1/2}}%
\int_{-\infty}^{+\infty}d_{2}^{n}p\,(\bar{u}_{R,\bar{L}})_{\ominus_{R}%
p,m}(x^{A},t=0))\overset{x|p}{\odot}_{\hspace{-0.01in}L}(C_{2})_{\kappa^{-1}%
p}(t)\nonumber\\
&  =\frac{\kappa^{-n}}{(\text{vol}_{2})^{1/2}}\big \langle(u_{R,\bar{L}%
})_{\ominus_{\bar{L}}p,m}(x^{A},t=0),(C_{2})_{\kappa^{-1}p}%
(t)\big \rangle_{2,p},
\end{align}
and%
\begin{align}
(\Psi_{1}^{\ast})_{m}(x^{i})  &  =(\text{vol}_{1})^{1/2}\int_{-\infty
}^{+\infty}d_{1}^{n}p\,(u_{\bar{R},L})_{p,m}(x^{A},t=0)\overset{p}%
{\circledast}(C_{1}^{\ast})_{\kappa^{-1}p}(t)\nonumber\\
&  =(\text{vol}_{1})^{1/2}\big \langle(\bar{u}_{\bar{R},L})_{p,m}%
(x^{A},t=0),(C_{1}^{\ast})_{\kappa^{-1}p}(t)\big \rangle_{1,p},\\[0.1in]
(\Psi_{2}^{\ast})_{m}^{\prime}(x^{i})  &  =(\text{vol}_{2})^{1/2}\int
_{-\infty}^{+\infty}d_{2}^{n}p\,(C_{2}^{\ast})_{\kappa p}^{\prime}%
(t)\overset{p}{\circledast}(\bar{u}_{R,\bar{L}})_{p,m}(x^{A},t=0)\nonumber\\
&  =(\text{vol}_{2})^{1/2}\big \langle(C_{2}^{\ast})_{\kappa p}^{\prime
}(t),(u_{R,\bar{L}})_{p,m}(x^{A},t=0)\big \rangle_{2,p}^{\prime}.
\label{EntInt4}%
\end{align}
Notice that throughout this section we take the convention that wave functions
and expansion coefficients referring to the interaction picture are written in
capital letters.

The relationship between wave functions of the interaction picture and those
of the Schr\"{o}dinger picture can be expressed as%
\begin{align}
(\Psi_{1})_{m}^{\prime}(x^{i})  &  =\exp(\text{i}q^{-\zeta}tH_{0})\overset
{x}{\triangleright}(\psi_{1})_{m}^{\prime}(x^{i}),\nonumber\\
(\Psi_{1}^{\ast})_{m}(x^{i})  &  =\exp(\text{i}tH_{0})\overset{x}%
{\triangleright}(\psi_{1}^{\ast})_{m}(x^{i}), \label{IntSch1}%
\end{align}
and%
\begin{align}
(\Psi_{2})_{m}(x^{i})  &  =(\psi_{2})_{m}(x^{i})\overset{x}{\triangleleft}%
\exp(-\text{i}q^{\zeta}H_{0}t),\nonumber\\
(\Psi_{2}^{\ast})_{m}^{\prime}(x^{i})  &  =(\psi_{2}^{\ast})_{m}^{\prime
}(x^{i})\overset{x}{\triangleleft}\exp(-\text{i}H_{0}t). \label{IntSch2}%
\end{align}
Let us note that in the above relations the operators on the right-hand side
remove the time-dependence resulting from the free-particle Hamiltonian
$H_{0}.$

With the very same reasonings as in the undeformed case (see for example Ref.
\cite{Sak94})\ it follows from the identities in (\ref{IntSch1}) and
(\ref{IntSch2}) that we have%
\begin{align}
\text{i}\partial_{0}\overset{t}{\triangleright}(\Psi_{1})_{m}^{\prime}(x^{i})
&  =(V_{1})_{I}^{\prime}\overset{x}{\triangleright}(\Psi_{1})_{m}^{\prime
}(x^{i}),\nonumber\\
\text{i}\partial_{0}\overset{t}{\triangleright}(\Psi_{1}^{\ast})_{m}(x^{i})
&  =(V_{1}^{\ast})_{I}\overset{x}{\triangleright}(\Psi_{1}^{\ast})_{m}(x^{i}),
\label{DifInt1}%
\end{align}
and
\begin{align}
(\Psi_{2})_{m}(x^{i})\overset{t}{\triangleleft}(\text{i}\hat{\partial}_{0})
&  =(\Psi_{2})_{m}(x^{i})\overset{x}{\triangleleft}(V_{2})_{I},\nonumber\\
(\Psi_{2}^{\ast})_{m}^{\prime}(x^{i})\overset{t}{\triangleleft}(\text{i}%
\hat{\partial}_{0})  &  =(\Psi_{2}^{\ast})_{m}^{\prime}(x^{i})\overset
{x}{\triangleleft}(V_{2}^{\ast})_{I}, \label{DifInt2}%
\end{align}
where
\begin{align}
(V_{1})_{I}^{\prime}  &  =\exp(\text{i}tq^{-\zeta}H_{0})V\exp(-\text{i}%
tq^{-\zeta}H_{0}),\nonumber\\
(V_{2})_{I}  &  =\exp(-\text{i}tq^{\zeta}H_{0})V\exp(\text{i}tq^{\zeta}H_{0}),
\end{align}
and%
\begin{align}
(V_{1}^{\ast})_{I}  &  =\exp(\text{i}tH_{0})V\exp(-\text{i}tH_{0}),\nonumber\\
(V_{2}^{\ast})_{I}^{\prime}  &  =\exp(-\text{i}tH_{0})V\exp(\text{i}tH_{0}).
\end{align}

To find solutions to the differential equations in (\ref{DifInt1}) and
(\ref{DifInt2}) we introduce time-evolution operators for the wave functions
of the interaction picture, i.e.
\begin{align}
(\Psi_{1})_{m}^{\prime}(x^{A},t)  &  =(U_{1})_{I}^{\prime}(t,\tilde
{t})\overset{x}{\triangleright}(\Psi_{1})_{m}^{\prime}(x^{A},\tilde
{t}),\nonumber\\
(\Psi_{1}^{\ast})_{m}(x^{A},t)  &  =(U_{1}^{\ast})_{I}(t,\tilde{t})\overset
{x}{\triangleright}(\Psi_{1}^{\ast})_{m}(x^{A},\tilde{t}), \label{DefUOpInt1}%
\end{align}
and%
\begin{align}
(\Psi_{2})_{m}(x^{A},t)  &  =(\Psi_{2})_{m}(x^{A},\tilde{t})\overset
{x}{\triangleleft}(U_{2})_{I}(t,\tilde{t}),\nonumber\\
(\Psi_{2}^{\ast})_{m}^{\prime}(x^{A},t)  &  =(\Psi_{2}^{\ast})_{m}^{\prime
}(x^{A},\tilde{t})\overset{x}{\triangleleft}(U_{2}^{\ast})_{I}^{\prime
}(t,\tilde{t}). \label{DefUOpInt2}%
\end{align}
Inserting these expressions into the differential equations in (\ref{DifInt1})
and (\ref{DifInt2}) yields differential equations for the time evolution
operators in the interaction picture:%
\begin{align}
\text{i}\partial_{0}\overset{t}{\triangleright}(U_{1})_{I}^{\prime}%
(t,\tilde{t})  &  =(V_{1})_{I}^{\prime}(t)\,(U_{1})_{I}^{\prime}(t,\tilde
{t}),\nonumber\\
\text{i}\partial_{0}\overset{t}{\triangleright}(U_{1}^{\ast})_{I}(t,\tilde
{t})  &  =(V_{1}^{\ast})_{I}(t)\,(U_{1}^{\ast})_{I}(t,\tilde{t}),
\label{SchGleInt1}%
\end{align}
and
\begin{align}
(U_{2})_{I}(t,\tilde{t})\overset{t}{\triangleleft}(\text{i}\hat{\partial}%
_{0})  &  =(U_{2})_{I}(t,\tilde{t})\,(V_{2})_{I}(t),\nonumber\\
(U_{2}^{\ast})_{I}^{\prime}(t,\tilde{t})\overset{t}{\triangleleft}%
(\text{i}\hat{\partial}_{0})  &  =(U_{2}^{\ast})_{I}^{\prime}(t,\tilde
{t})\,(V_{2}^{\ast})_{I}(t). \label{SchGleInt2}%
\end{align}

If we require%
\begin{equation}
(U_{1})_{I}^{\prime}(t,t)=(U_{1}^{\ast})_{I}(t,t)=(U_{2})_{I}(t,t)=(U_{2}%
^{\ast})_{I}^{\prime}(t,t)=1,
\end{equation}
the differential equations in (\ref{SchGleInt1}) and (\ref{SchGleInt2}) are
equivalent to the integral equations
\begin{align}
(U_{1})_{I}^{\prime}(t,\tilde{t})  &  =1-\text{i}\int_{\tilde{t}}%
^{t}dt^{\prime}\,(V_{1})_{I}^{\prime}(t^{\prime})\,(U_{1})_{I}^{\prime
}(t^{\prime},\tilde{t}),\nonumber\\
(U_{1}^{\ast})_{I}(t,\tilde{t})  &  =1-\text{i}\int_{\tilde{t}}^{t}dt^{\prime
}\,(V_{1}^{\ast})_{I}(t^{\prime})\,(U_{1}^{\ast})_{I}(t^{\prime},\tilde{t}),
\end{align}
and%
\begin{align*}
(U_{2})_{I}(t,\tilde{t})  &  =1+\text{i}\int_{t^{\prime}}^{t}dt^{\prime
}\,(U_{2})_{I}(t^{\prime},\tilde{t})\,(V_{2})_{I}(t^{\prime}),\\
(U_{2}^{\ast})_{I}^{\prime}(t,\tilde{t})  &  =1+\text{i}\int_{t^{\prime}}%
^{t}dt^{\prime}\,(U_{2}^{\ast})_{I}^{\prime}(t^{\prime},\tilde{t}%
)\,(V_{2}^{\ast})_{I}^{\prime}(t^{\prime}),
\end{align*}
By iteration we find as formal solutions%
\begin{align}
(U_{1})_{I}^{\prime}(t,\tilde{t})  &  =1+\sum_{n=1}^{\infty}\text{i}^{-n}%
\int_{\tilde{t}}^{t}dt_{1}\int_{\tilde{t}}^{t_{1}}dt_{2}\ldots\int_{\tilde{t}%
}^{t_{n-1}}dt_{n}\,(V_{1})_{I}^{\prime}(t_{1})\ldots(V_{1})_{I}^{\prime}%
(t_{n}),\nonumber\\
(U_{1}^{\ast})_{I}(t,\tilde{t})  &  =1+\sum_{n=1}^{\infty}\text{i}^{-n}%
\int_{\tilde{t}}^{t}dt_{1}\int_{\tilde{t}}^{t_{1}}dt_{2}\ldots\int_{\tilde{t}%
}^{t_{n-1}}dt_{n}\,(V_{1}^{\ast})_{I}(t_{1})\ldots(V_{1}^{\ast})_{I}(t_{n}),
\end{align}
and%
\begin{align}
(U_{2})_{I}(t,\tilde{t})  &  =1+\sum_{n=1}^{\infty}\text{i}^{n}\int
_{t^{\prime}}^{t}dt_{1}\int_{t^{\prime}}^{t_{1}}dt_{2}\ldots\int_{t^{\prime}%
}^{t_{n-1}}dt_{n}\,(V_{2})_{I}(t_{n})\ldots(V_{2})_{I}(t_{1}),\nonumber\\
(U_{2}^{\ast})_{I}^{\prime}(t,\tilde{t})  &  =1+\sum_{n=1}^{\infty}%
\text{i}^{n}\int_{t^{\prime}}^{t}dt_{1}\int_{t^{\prime}}^{t_{1}}dt_{2}%
\ldots\int_{t^{\prime}}^{t_{n-1}}dt_{n}\,(V_{2}^{\ast})_{I}(t_{n})\ldots
(V_{2}^{\ast})_{I}(t_{1}).
\end{align}

It remains to write down formulae that extract the coefficients from the
expansions in (\ref{EntInt1})-(\ref{EntInt4}). Recalling the results in Ref.
\cite{Qkin1} this task can be achieved by the formulae%
\begin{align}
(C_{1})_{p}^{\prime}(t)  &  =(\text{vol}_{1})^{1/2}\int\nolimits_{-\infty
}^{+\infty}d_{1}^{n}x\,(\Psi_{1})_{m}^{\prime}(x^{i})\overset{x}{\circledast
}(u_{\bar{R},L})_{p,m}(x^{A},t=0)\nonumber\\
&  =(\text{vol}_{1})^{1/2}\big \langle(\Psi_{1})_{m}^{\prime}(x^{i}),(\bar
{u}_{\bar{R},L})_{p,m}(x^{A},t=0)\big \rangle_{1,x}^{\prime}%
,\label{ExpCoeInt1}\\[0.1in]
(C_{2})_{p}(t)  &  =(\text{vol}_{2})^{1/2}\int\nolimits_{-\infty}^{+\infty
}d_{2}^{n}x\,(\bar{u}_{R,\bar{L}})_{p,m}(x^{A},t=0)\overset{x}{\circledast
}(\Psi_{2})_{m}(x^{i})\nonumber\\
&  =(\text{vol}_{2})^{1/2}\big \langle(u_{R,\bar{L}})_{p,m}(x^{A}%
,t=0),(\Psi_{2})_{m}(x^{i})\big \rangle_{2,x},
\end{align}
and%
\begin{align}
(C_{1}^{\ast})_{p}(t)  &  =(\text{vol}_{1})^{-1/2}\int\nolimits_{-\infty
}^{+\infty}d_{1}^{n}x\,(u_{\bar{R},L})_{\ominus_{\bar{R}}p,m}(x^{A}%
,t=0)\overset{p|x}{\odot}_{\hspace{-0.01in}\bar{L}}(\Psi_{1}^{\ast})_{m}%
(x^{i})\nonumber\\[0.1in]
&  =(\text{vol}_{1})^{-1/2}\big \langle(\bar{u}_{\bar{R},L})_{\ominus_{L}%
p,m}(x^{A},t=0),(\Psi_{1}^{\ast})_{m}(x^{i})\big \rangle_{1,x},\\
(C_{2}^{\ast})_{p}^{\prime}(t)  &  =(\text{vol}_{2})^{-1/2}\int
\nolimits_{-\infty}^{+\infty}d_{2}^{n}x\,(\Psi_{2}^{\ast})_{m}^{\prime}%
(x^{i})\overset{x|p}{\odot}_{\hspace{-0.01in}\bar{R}}(\bar{u}_{\bar{R}%
,L})_{\ominus_{\bar{L}}p,m}(x^{A},t=0)\nonumber\\
&  =(\text{vol}_{2})^{-1/2}\big \langle(\Psi_{2}^{\ast})_{m}^{\prime}%
(x^{i}),(u_{\bar{R},L})_{\ominus_{R}p,m}(x^{A},t=0)\big \rangle_{2,x}.
\label{ExpCoeInt2}%
\end{align}
These equalities are a direct consequence of the completeness relations for
momentum eigenfunctions. In this sense, they can be viewed as a kind of
inverse Fourier transformation.

As already mentioned the expansion coefficients in (\ref{ExpCoeInt1}%
)-(\ref{ExpCoeInt2}) are nothing other than transition amplitudes which
contain information about finding a system in a certain momentum eigenstate.
In what follows we would like to adapt these ideas in a way suitable for
describing scattering processes.

Towards this end let us first recall that in the interaction picture
free-particle wave functions are time-independent. Thus, they are linked
to\ the free-particle wave functions of the Schr\"{o}dinger picture via the
relations%
\begin{align}
(\Phi_{1})_{m}^{\prime}(x^{A})  &  =(\phi_{1})_{m}^{\prime}(x^{A}%
,t_{x}=0),\nonumber\\
(\Phi_{2})_{m}(x^{A})  &  =(\phi_{2})_{m}(x^{A},t_{x}=0),\\[0.1in]
(\Phi_{1}^{\ast})_{m}(x^{A})  &  =(\phi_{1}^{\ast})_{m}(x^{A},t_{x}%
=0),\nonumber\\
(\Phi_{2}^{\ast})_{m}^{\prime}(x^{A})  &  =(\phi_{2}^{\ast})_{m}^{\prime
}(x^{A},t_{x}=0).
\end{align}

In scattering processes one requires for the particles to be free in the
remote past or remote future. For wave functions of the interaction picture
these conditions read%
\begin{align}
\lim_{t\rightarrow+\infty}(\Psi_{2})_{m^{-}}(x^{A},t)  &  =(\Phi_{2}%
)_{m}(x^{A}),\nonumber\\
\lim_{t\rightarrow-\infty}(\Psi_{1})_{m^{+}}^{\prime}(x^{A},t)  &  =(\Phi
_{1})_{m}^{\prime}(x^{A}),\label{BouCon1}\\[0.1in]
\lim_{t\rightarrow-\infty}(\Psi_{1}^{\ast})_{m^{+}}(x^{A},t)  &  =(\Phi
_{1}^{\ast})_{m}(x^{A}),\nonumber\\
\lim_{t\rightarrow+\infty}(\Psi_{2}^{\ast})_{m^{-}}^{\prime}(x^{A},t)  &
=(\Phi_{2}^{\ast})_{m}^{\prime}(x^{A}). \label{BouCon2}%
\end{align}

The amplitudes that determine the probability for detecting a certain free
particle state are given by%
\begin{align}
(S_{2})_{-}(\Phi,\Psi)  &  =\lim_{t\rightarrow-\infty}\big \langle(\Phi
_{2}^{\ast})_{m}(x^{A}),(\Psi_{2})_{m^{-}}(x^{B},t)\big \rangle_{2,x}%
\nonumber\\
&  =\big \langle(\Phi_{2}^{\ast})_{m}(x^{A}),(\Phi_{2})_{m}(x^{B})\overset
{x}{\triangleleft}(U_{2})_{I}(-\infty,+\infty)\big \rangle_{2,x}%
,\label{SMaInt1}\\[0.1in]
(S_{1}^{\ast})_{+}(\Phi,\Psi)  &  =\lim_{t\rightarrow+\infty}\big \langle(\Phi
_{1})_{m}(x^{A}),(\Psi_{1}^{\ast})_{m^{+}}(x^{B},t)\big \rangle_{1,x}%
\nonumber\\
&  =\big \langle(\Phi_{1})_{m}(x^{A}),(U_{1}^{\ast})_{I}(+\infty
,-\infty)\overset{x}{\triangleright}(\Phi_{1}^{\ast})_{m}(x^{B}%
,t)\big \rangle_{1,x},
\end{align}
and%
\begin{align}
(S_{1})_{+}^{\prime}(\Phi,\Psi)  &  =\lim_{t\rightarrow+\infty}%
\big \langle(\Psi_{1})_{m^{+}}^{\prime}(x^{A},t),(\Phi_{1}^{\ast})_{m}%
^{\prime}(x^{B})\big \rangle_{1,x}^{\prime}\nonumber\\
&  =\big \langle(U_{1})_{I}^{\prime}(+\infty,-\infty)\overset{x}%
{\triangleright}(\Phi_{1})_{m}^{\prime}(x^{A}),(\Phi_{1}^{\ast})_{m}^{\prime
}(x^{B})\big \rangle_{1,x}^{\prime},\\[0.1in]
(S_{2}^{\ast})_{-}^{\prime}(\Phi,\Psi)  &  =\lim_{t\rightarrow-\infty
}\big \langle(\Psi_{2}^{\ast})_{m^{-}}^{\prime}(x^{A},t),(\Phi_{2}%
)_{m}^{\prime}(x^{B})\big \rangle_{2,x}^{\prime}\nonumber\\
&  =\big \langle(\Phi_{2}^{\ast})_{m}^{\prime}(x^{A})\overset{x}%
{\triangleleft}(U_{2}^{\ast})_{I}^{\prime}(-\infty,+\infty),(\Phi_{2}%
)_{m}^{\prime}(x^{B})\big \rangle_{2,x}^{\prime}. \label{SMaInt4}%
\end{align}
The second equality in each of the above equations follows from the relations
in (\ref{DefUOpInt1}) and (\ref{DefUOpInt2}) together with the boundary
conditions in (\ref{BouCon1}) and (\ref{BouCon2}).

\section{Conclusion\label{SecCon}}

Let us make some comments on the results of our examinations about a q-analog
of non-relativistic Schr\"{o}dinger theory.

In part I of the present paper we presented a mathematical and physical
framework that can lead to theories with the attractive feature that space is
discretized but time is not. Based on these ideas we worked out the basics of
a non-relativistic Schr\"{o}dinger theory on q-deformed quantum spaces as the
braided line and the three-dimensional q-deformed Euclidean space. This was
done in part II and III of our article.

Contrary to many other approaches for introducing a lattice-like structure in
physics our theory does not suffer from the absence of important space-time
symmetries, such as rotational or translational symmetry. For this reason, it
can be developed along the same line of reasonings as its undeformed
counterpart. In view of this observations our q-deformed version of
Schr\"{o}dinger theory seems to be of the same value as the classical
non-relativistic Schr\"{o}dinger theory, which we regain from our results when
the deformation parameter $q$ tends to 1.

However, q-deformation of physical theories gives rise to some more structure.
For example, we saw that q-deformation requires to distinguish different
geometries, which become identical in the undeformed case. To get a complete
description of physical phenomena one cannot restrict attention to one such
geometry. This circumstance makes things more difficult, but it can lead to
new physical phenomena.

On the other hand a special quality of our approach is that it generalizes and
extends a well established theory in a way that the relationship to the
undeformed limit is rather clear. So to speak, the classical theory can be
seen as an approximation or simplification of a more detailed description. In
this manner, our q-deformed version of non-relativistic Schr\"{o}dinger theory
can be a useful step in examining and understanding the implications of
q-deformation on quantum mechanics, since the treatment of more realistic
space-time structures like the q-deformed Minkowski space \cite{CSSW90, PW90,
SWZ91, Maj91, LWW97} is very awkward (for other deformations of space-time see
also Refs. \cite{Lu92, Cas93, Dob94, DFR95, ChDe95, ChKu04, Koch04}%
).\vspace{0.16in}

\noindent\textbf{Acknowledgements}

First of all I am very grateful to Eberhard Zeidler for very interesting and
useful discussions, special interest in my work and financial support.
Furthermore, I would like to thank Alexander Schmidt for useful discussions
and his steady support. Finally, I thank Dieter L\"{u}st for kind hospitality.

\end{document}